\documentclass[twocolumn,amssymb, nobibnotes, aps, prc,superscriptaddress]{revtex4-1}
\usepackage[utf8]{inputenc}
\usepackage{graphicx}
\usepackage{dcolumn}
\usepackage{bm}
\usepackage{epstopdf}
\usepackage{appendix}

\begin{document}

\title{Interpretation of normal-deformed bands in $^{167}$Lu}
\author{Azam Kardan}
\email{aakardan@du.ac.ir (Corresponding author)}
\affiliation{School of Physics, Damghan University, P.O. Box 36716-41167, Damghan, Iran}
\author{Ingemar Ragnarsson}
\email{ingemar.ragnarsson@matfys.lth.se}
\affiliation{Division of Mathematical Physics, LTH, Lund University, P.O. Box 118, SE-221 00 Lund, Sweden}
\author{B. Gillis Carlsson}
\affiliation{Division of Mathematical Physics, LTH, Lund University, P.O. Box 118, SE-221 00 Lund, Sweden}
\author{Hai-Liang Ma}
\affiliation{China Institute of Atomic Energy, P.O. Box 275-10, Beijing 102413, China}

\date{\today}

\begin{abstract}
With the aim to get a general understanding of rotational bands in the deformed
rare-earth region or in deformed nuclei in general, the observed normal-deformed rotational
structures in $^{167}$Lu are interpreted within the unpaired and paired
cranked Nilsson-Strutinsky formalisms, CNS and CNSB.
Particular attention is devoted to the band crossings.
For this nucleus with the Fermi surface high up in the $h_{11/2}$ shell,
we conclude that except for the paired AB and BC crossings in configurations with
an even and odd number of $i_{13/2}$ neutrons, respectively, the observed band
crossings can be understood within the unpaired formalism. Especially,
it means that above the AB and BC crossings, the evolution with spin is 
described as a gradual alignment of the spin
vectors of the particles outside closed shells. Consequently, the configurations
can be characterized by the number of particles occupying open $j$-shells or groups of
$j$-shells. In the present study, we revise the interpretation of some experimental
bands and also the nature of the crossings while some previous
configuration assignments are confirmed.
\end{abstract}
\maketitle

\section{Introduction}
In the analysis of nuclear high-spin states, much emphasis has been
put on band crossings caused by alignments of high-$j$ particles.
Such band crossings are formed when
the Coriolis and centrifugal forces overcome the pairing forces which
try to keep the particles in time-reversed orbitals. Specifically, after
the discovery of the backbending 
phenomenon almost 50 years ago \cite{Joh72}, it was soon
realized that it was caused by this type of band crossings \cite{Ste75}.
Subsequently, 
with the assumption that it is a reasonable approximation to 
assume that the deformation and the 
pairing gap stay constant over the crossing region, a simple and
illustrative formalism \cite{Boh78,Ben79}, the Cranked Shell Model 
(CSM) was developed in 
Copenhagen.
This 
formalism has been very successful to explain e.g.\ the details of the
first crossing in the rare-earth region, caused by the $i_{13/2}$ neutrons, 
and generally also to describe
the second crossing in these nuclei caused by the $h_{11/2}$ protons, see e.g.\ Refs.\ \cite{Rie81,Voi83}.

When going to higher spin values, the Coriolis and centrifugal forces
will lead to shape changes and configuration changes not accounted for
by the CSM in its standard form. With the purpose to treat these
features in detail the so called Cranked Nilsson Strutinsky (CNS) formalism
\cite{beng85}
was developed. The main purpose of this formalism is to describe high
and very high spin states, where pairing correlations are
assumed to be of minor importance and are therefore neglected. 
In both the CSM and CNS formalisms, diabatic configurations are
followed through band-crossings, where computer codes are used
in the CNS formalism to do this in a systematic way over the
entire deformation space. With some further developments of the
formalism \cite{Rag95, carl06}, it is now 
possible to fix
configurations in a detailed way based on the occupation
of ${\cal N}$-shells, and of $j$-shells or groups of $j$-shells within 
each ${\cal N}$-shell. 

With the background of the CNS model,
the Ultimate cranker model was developed 
in the late 1980's 
\cite{beng89, beng90}. 
%
%
%
In this model, it is in principle possible
to follow diabatic 
configurations and carry out self-consistent
calculations in both pairing space and deformation space.
However, this requires heavy computer calculations which were
very difficult not to say impossible to carry through in the early
years of the model. 
Even so, already at the introduction of the Ultimate cranker, attempts
were made to follow the deformation changes and the decrease of
pairing in the rotational bands in $^{158}$Er when they approach termination
in the $I > 40 \hbar$ spin range \cite{beng90}. 
Furthermore, for the superdeformed bands
in $^{143}$Eu, the details of the decrease of pairing with increasing angular
momentum has been investigated \cite{Axe02}.

The mentioned studies on $^{158}$Er and $^{143}$Eu are however rather
exceptions and the Ultimate cranker has
mainly been used to apply the CSM formalism. Configurations are then labeled
by the number of aligned particles, see e.g.\ \cite{rou15} and
references therein. In our view, this
becomes questionable when going beyond the first and maybe second 
band-crossing because pairing is then severely quenched and it 
is not meaningful 
to make a division into spin vectors which are aligned
and those which are not aligned. Indeed, when the CSM model was
introduced \cite{Boh78,Ben79} it was clearly stated that it was mainly intended for
the understanding of rotating deformed nuclei up to an angular
momentum of about $30 \hbar$.


The possibility to get a detailed understanding of very high spin
states has been demonstrated in the CNS model but mainly on nuclei where
it has been possible to follow configurations 
which reach or come close to termination and on superdeformed states.
Thus, examples of successful studies of high-spin level schemes
within the CNS formalism
are 
spread over the nuclear
periodic table, e.g.\ $^{20}$Ne \cite{Rag81} and $^{38}$Ar \cite{Sve01} where 
strongly deformed bands have 
been followed to termination,  $^{62}$Zn where the excitation of p-h excitations
can be followed from the ground state to superdeformation \cite{ge12}, 
 $^{74}$Kr \cite{Val05} where a band is followed to an $I_{max}$ state which is still collective,   
$^{109,110}$Sb \cite{Rag95,Afa99} with a detailed prediction of smooth terminating states,  
$^{113}$I \cite{Sta01} where experiment is coming close to a very high spin state 
which terminates high above yrast, the superdeformed bands in the $^{152}$Dy region
\cite{Rag93} and 
 $^{157,158}$Er \cite{Eva06, Sim94} with a detailed understanding not only of
the high-spin terminations but, for  $^{157}$Er, also of the feeding of the
terminating states.

 A few years ago,
 the Ultimate Cranker formalism was updated so that it is now
 straightforward to make calculations in a mesh covering both the deformation space and
 pairing space, the Cranked Nilsson-Strutinsky-Bogoliubov (CNSB) formalism
 \cite{carl08,ma14}. Combined
 with the CNS formalism and its possibility to label configurations in detail,
 this leads to a very powerful formalism to analyze high-spin states. Here we will
 test this formalism on a nucleus in the middle of the deformed rare-earth region,
 namely on $^{167}$Lu where one of the most extensive level schemes in this region of nuclei
 has recently been published, see Ref.\ \cite{rou15}. The high-spin states in Lu nuclei are particularly interesting since
it is in these isotopes that one has observed \cite{bri05,Jen02,Sch03,amro03} the nuclear wobbling
excitation. Observation of the wobbling mode gives firm evidence that
nuclei are triaxial and shows to what degree nuclei can be seen as
rigid triaxial bodies. While several articles have been devoted solely to
the study of the wobbling mode, ours is the first detailed study of the
remaining states and shows how well theory can describe the states of
this nucleus. The nucleus $^{167}$Lu is of special interest because
it is the only case where an interaction has been observed at high spin between 
the normal-deformed states and the wobbling band. Note also that,
mainly because of the wobbling excitations identified in
this region of nuclei, extensive level schemes have been published for
many neighboring nuclei, see e.g.\ Refs.\ \cite{bri06,jen04,bri07,sch04,mar13,yad09}.

In Ref.\ \cite{rou15}, the standard
procedure was followed, assuming that, with increasing angular momentum, the different 
rotational bands can be characterized by an increasing number of particles that have
their spin vectors aligned. As mentioned above we do not find such a description
meaningful for very high spin states, say $I>30$ in the deformed rare earth region.
Indeed, our calculations below show that such high-spin bands
in $^{167}$Lu are well understood from configurations with a fixed number of particles
in different (groups of) $j$-shells where the spin vectors of the valence 
particles become gradually
more and more aligned with increasing rotational frequency. We believe that these results are of a more
general nature, where similar conclusions can be
drawn from our previous studies of $^{161}$Lu \cite{ma14} which is rather a transitional
nucleus in the outskirts of the deformed rare earth region.
 

The observed bands of $^{167}$Lu are analyzed in Sec.~\ref{labeling} where special 
emphasis is put on interactions
between different bands. Our theoretical models, CNS and CNSB, are briefly described in
Sec.~\ref{model}. The observed bands are then analyzed in these models in Sec.~\ref{analyze} and 
configurations are assigned to the different bands. Finally, our results are summarized
in Sec.~\ref{summ} with some brief suggestions for future experiments. In an appendix, we
discuss how the alignment at band crossings should be defined and how this alignment can be
determined for observed rotational bands.

\section{Experimental bands and their labeling}
\label{labeling}
\begin{figure}
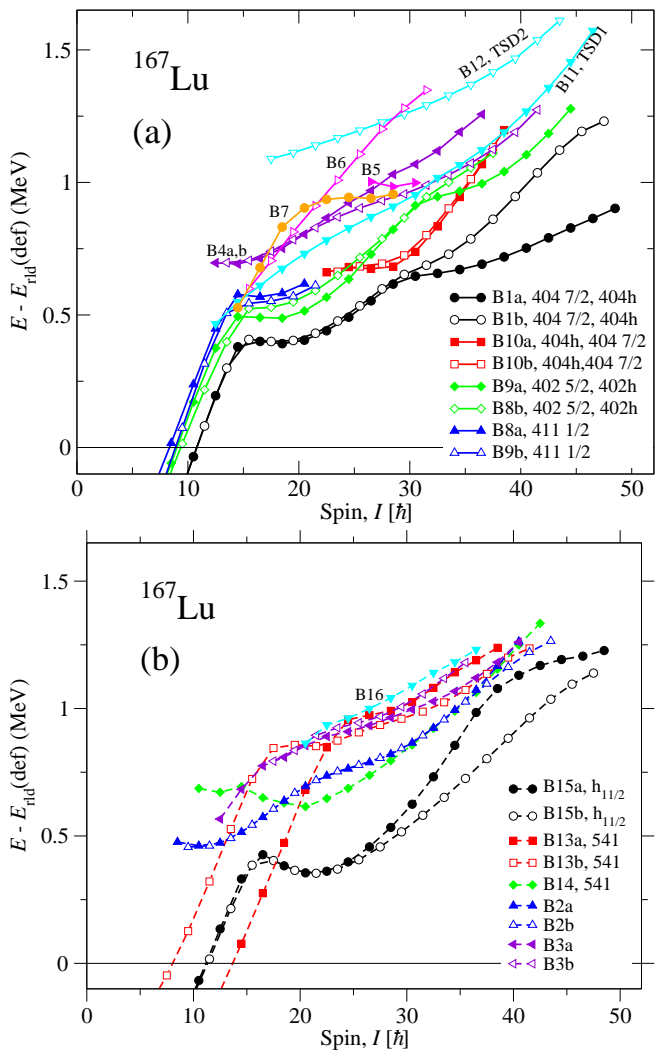

 \begin{centering}
\includegraphics[clip=true,width=0.48\textwidth,trim=0 0 0 0]{erefr2-rw-p.eps}
\includegraphics[clip=true,width=0.48\textwidth,trim=0 0 0 0]{erefr2-rw-n.eps}
  \caption{\label{exp}  Energies of the experimental (a) positive and (b) negative 
parity bands in $^{167}$Lu. The bands are labeled by the numbers used in
Ref.\cite{rou15} but for some bands we define different labels which are also
given and which  are used in our interpretation. For the proton bands based
on the ${\cal N}=4$ orbitals, see also Fig.\ \ref{exp4} below. The negative
parity orbitals interpreted as built on an $h_{11/2}$ or on the [541]1/2 $hf$
orbital are labeled as $h_{11/2}$ and 541, respectively. The positive
(negative) parity bands are drawn by full (dashed) lines while closed (open)
symbols are used for signature $\alpha = 1/2$ ($-1/2$) bands. This convention
is used for all figures.}
 \end{centering}
\end{figure}

The experimental positive and negative parity bands in $^{167}$Lu are
drawn in Fig.\ \ref{exp} (a) and (b), respectively. They are taken from
Ref.\cite{rou15}, where they are all labeled with a number and
in most cases also with one or several letters referring to
which orbitals are considered aligned.
However, as discussed below, in some cases, we define 
bands in a different way; e.g.
when two bands interact we try to form smooth undisturbed crossing bands.  
Thus, we have added labels on some of the bands
referring to our interpretation. 
For such bands we will mainly use these 
labels, defined in Fig.\ \ref{exp}, when comparing with calculations, while
for other bands, we will use the original labels of Ref.\cite{rou15}.
In our labels, the letter $a$ is used
for signature $\alpha=1/2$ bands and $b$ for $\alpha=-1/2$
bands. 

Consider first the positive parity bands. As discussed in
connection with band 10 in Ref.\cite{rou15}, there is a crossing
between bands 1 and 10 at $I \approx 30$. Furthermore, there appears to be
a crossing between band 9a and the TSD1 band, also at $I \approx 30$. These
crossings will be discussed in detail below and we will conclude that
the crossing bands at the yrast line should be assigned as [404]7/2
and a three-quasiparticle configuration with 7 $h_{11/2}$ protons and
5 $i_{13/2}$ neutrons (labeled as 404h). 
Furthermore, in agreement with Ref.\cite{rou15}, the 
band labeled 4 is interpreted as a three-quasiparticle configuration. 

For negative parity, the low spin range of band 15 is built on the
$h_{11/2}$ orbital where it turns out to be somewhat questionable if it
is the [523]7/2 or the [514]9/2 orbital which is involved. Thus, we 
will refer to it as the $h_{11/2}$ band.
The bands 13 and 14 appear to be associated with the
[541]1/2 orbital. 
Thus, we will sometimes refer to these bands as the 541 bands.
Bands
2 and 3 are assigned as 3-quasiparticle bands in agreement with
Ref.\cite{rou15}.

 Bands 6 and 7 are rather short but we will give some comments on them
in Sec.~\ref{n4} while band 5 is too short to make it meaningful to
suggest any assignment. The spin and parity of band 16 is uncertain. 
However, with the values suggested in Ref.\cite{rou15} and used in
Fig.\ \ref{exp}(b), the $E$ vs. $I$ curve of the band becomes similar to that
of bands 13 and 3. This indicates that the values in Ref.\cite{rou15}
are correct because with other values, the band would become too
different compared with the other observed bands. We will give some
short comments on this band in Sec.~\ref{541}.

Bands 11 and 12 are
generally labeled as triaxial strongly deformed, TSD1 and TSD2, with
similar bands observed in $^{163,165}$Lu \cite{Jen02,Sch03}. The structure of the TSD1 bands
is accepted since long, see Refs.\ 
\cite{Sch95,Ben04,Car07,kar12}.
The TSD2 bands are understood as a wobbling excitation built on the TSD1
bands, see \cite{Ham03,Car07,Shi08,Sug10,Alm11,Fra14}. 
However, using similar methods as employed
here, it has recently been pointed out 
\cite{Rag17} that there are some problems with the interpretation
that these bands are formed in a strongly deformed triaxial minimum.
Here, we will add some comments on how the TSD1 band interacts with
the normal-deformed bands but we have nothing to add on 
the interpretation of these `TSD' bands.

\begin{figure}
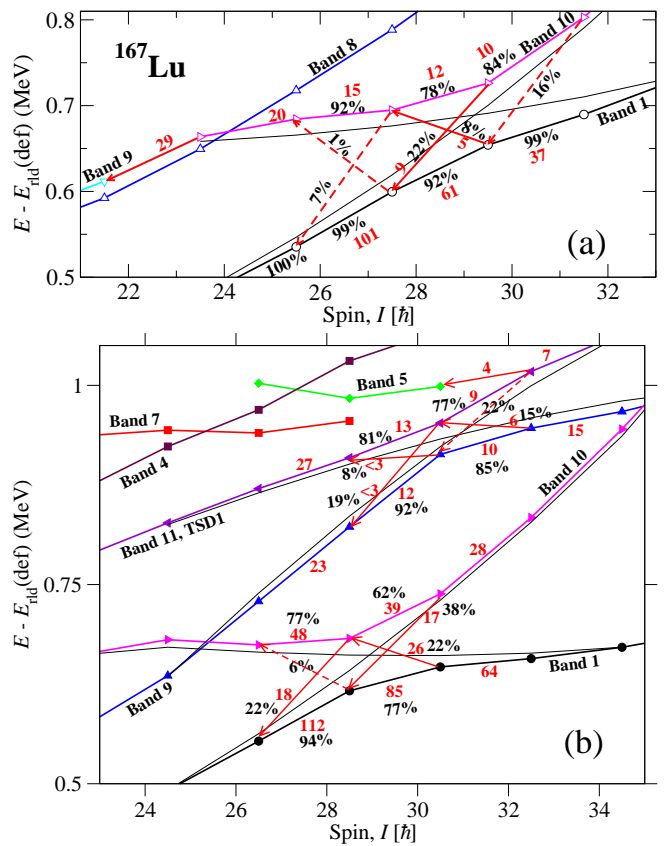

 \begin{centering}
\includegraphics[clip=true,width=0.48\textwidth,trim=0 0 0 0,angle=0]{Lu167-rw-ags-p-4.eps}
\includegraphics[clip=true,width=0.48\textwidth,trim=0 0 0 0,angle=0]{eref-pp-trans.eps}
  \caption{\label{exp1} Blown ups of Fig.\ref{exp}(a) in the spin
    region where the transitions are observed between (a) 
    bands 10 and 1 for $\alpha=-1/2$ and
    (b) bands 10 and 1, and bands 9 and 11 for $\alpha=+1/2$. Thin lines
are used for crossing bands fitted in two-band-mixing calculations,
see Sec.~\ref{undis}. Observed
    transitions are shown by full line red arrows while dashed arrows
    are used for
    transitions that are not observed. Experimental
    intensities are written in red while calculated branching ratios
    are given in percentage by black numbers.
}
 \end{centering}
\end{figure}

\subsection{Band crossings - smooth undisturbed bands}
\label{undis}
A closer study of the energy level scheme in Fig.\ \ref{exp}(a) 
indicates an interesting sequence of interactions
between the positive parity structures. Observed transitions which
link the bands are shown by arrows in Figs.\ \ref{exp1}(a) and (b), which are 
blow ups of Fig.\ \ref{exp}(a) for signature $\alpha = - 1/2$ and $1/2$,
respectively, in the spin region where the transitions are
observed.  

Specifically, both the odd and the even spin sequences
of bands 1 and 10 interact around $I=30$ where they come close
together. It suggests that bands 1 and 10 cross and exchange character
at $I \approx 30$.
Thus, in a similar way as was previously done for $^{76}$Rb \cite{Wad11},
we have performed a two-band-mixing calculation to find out if this
scenario is consistent with the observed bands. In this
calculation, two smooth unperturbed bands are parameterized
with a constant moment of inertia. They interact with strengths which
are assumed to be constant over the spin range $I = 23.5–33.5$
($I=24.5-34.5$) for the $\alpha=-1/2$ ($\alpha=1/2$) bands.  With a
least-square fit, all observed states in this spin range are
reproduced within $\pm7$ ($\pm5$) keV with an interaction matrix
element of 38 keV (30 keV). As seen in Fig.\ \ref{exp1} the two
undisturbed bands cross at $I \approx 29$ for both signatures.
The interaction between bands 1 and 10
is also discussed in Ref.\cite{rou15} where, in general
agreement with the present fit, a maximum 
interaction strength of $|V_{max}|=33$ keV is
deduced.  

Branching ratios for transitions between bands 1 and 10 are also 
calculated in the present two-band mixing model,
assuming the same transition strengths, B(E2)’s, within the smooth 
unperturbed bands and no transition probabilities connecting these bands.
These calculations show a reasonable agreement with
experiment in that the observed bands follow the strongest calculated
branching ratios, see Fig.\ref{exp1}(a) (Fig.\ref{exp1}(b)) for
$\alpha=-1/2$ ($\alpha=1/2$) bands.  Red numbers show the
experimental relative intensities while the calculated branching
between in-band and out-of-band transitions are shown in percentage
by black numbers.  For the $\alpha=-1/2$
bands in Fig.\ref{exp1}(a), the connecting transitions with a calculated
branching ratio of $22\%$ and $8\%$ at $I=29.5$ between the two bands
are observed. However, a transition
with a calculated $16\%$ branching ratio from the $I=31.5$ state
of band 10 to the
$I=29.5$ of band 1 is not seen in the experiment.
For $\alpha=1/2$ in Fig.\ref{exp1}(b), the connecting transitions with a
predicted $38\%$ and $22\%$ branching ratio are observed whilst the one
with a predicted $6\%$ branch is not observed.


Band 9a and the band which is assigned as TSD1, band 11, interact at
$I \approx 30$ in a similar way as bands 1 and 10, see 
Fig.\ref{exp1}(b).  Band mixing calculations were performed also for these
bands in the spin range $I=26.5-36.5$. 
With  the moment of inertia parameterized with a linear dependence on $I$, 
it was possible to reproduce all observed states in the studied spin range
within $\pm10$ keV, using   an interaction matrix element of 24 keV.
As seen in
Fig.\ref{exp1}(b), the band mixing calculation predicts a transition
with a $22\%$ branching ratio from the $I=32.5$ state of band 11 to
the $I=30.5$ state of band 9a which is not observed while the
transitions with a lower predicted branching ratio, $15\%$, $19\%$ and $8\%$ are
observed.
We note that this crossing is disturbed by other bands, bands 5 and 7,
which are only observed below the crossing. Therefore, the assumption
of a two-band crossing is questionable in this case. It is clear that
the bands interact but it is clearly more questionable if the two
bands `cross' and exchange character around $I=30$. Thus, in our
interpretation below, we will not treat these bands as crossing.
Furthermore, we will not try to interpret
band 11 (the TSD band) which has been discussed
in Ref.\ \cite{Rag17}.

\subsection{Bands based on ${\cal N}=4$ band heads}
\begin{figure*}
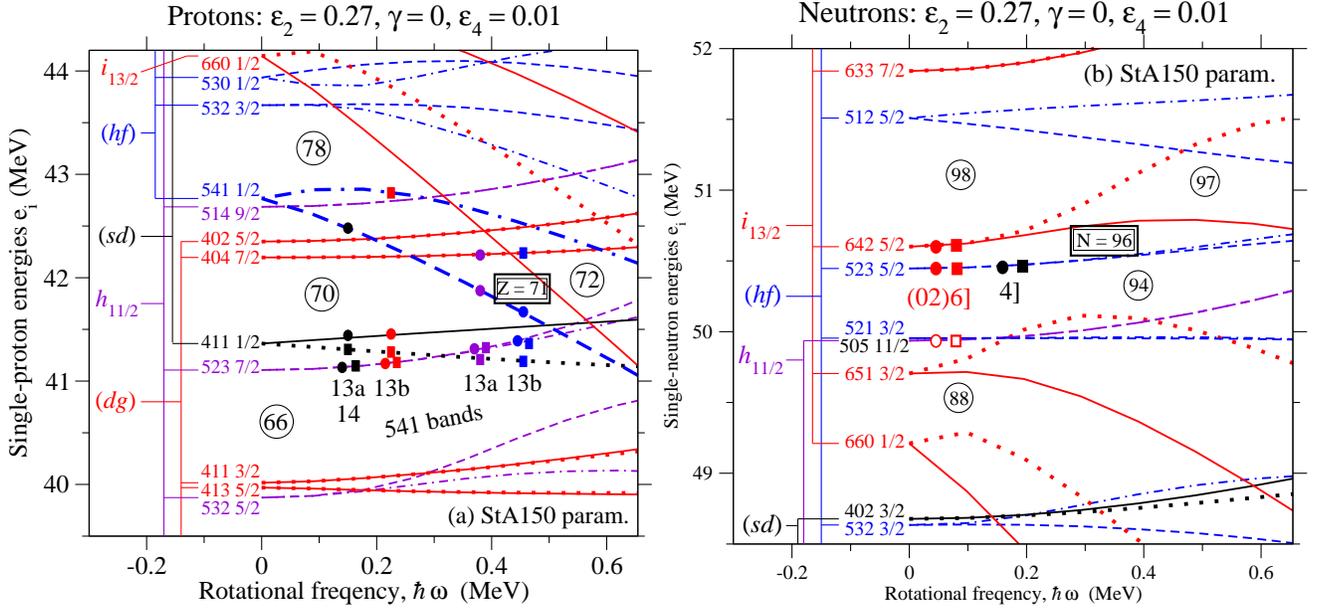

 \begin{centering}
\includegraphics[clip=true,width=0.48\textwidth,trim=0 0 0 0]{sppr5-e27g0.eps}
\includegraphics[clip=true,width=0.48\textwidth,trim=0 0 0 0]{spnr5-e27g0.eps}
  \caption{\label{sp} Single-routhian orbitals at an average deformation
calculated for the normal-deformed bands in $^{167}$Lu, 
$(\varepsilon_2,\gamma,\varepsilon_4) \approx (0.27,0^{\circ},0.01)$ for (a) protons and (b) neutrons. 
Dashed lines are used for negative parity
orbitals while dots are used for signature $\alpha = -1/2$. 
It is indicated how the proton orbitals can be classified as $dg$ ($d_{5/2}g_{7/2}$), 
$sd$ ($s_{1/2}d_{3/2}$), $h_{11/2}$ and $hf$ ($h_{9/2}f_{7/2}$), where it is not always possible
to distinguish between the low-$j$ ${\cal N}=4$ $dg$ and $sd$ orbitals. The neutron 
orbitals are classified as $hf$ ($h_{9/2}f_{7/2}$), $i_{13/2}$ and $h_{11/2}$.
The filling of the orbitals above the $Z=66$ and $N=94$ 
gaps according to our interpretation 
is illustrated for the 541 bands where circles are used for signature
$\alpha = 1/2$ and squares for $\alpha = -1/2$. At low spin for the 13a and 13b
configurations, the proton configurations illustrated at low frequencies are
combined with the (02)6] neutron configuration while at high spin, the proton
configurations illustrated at high frequencies are combined with the 4] neutron
configuration. Band 14 is built from the proton configuration to the left combined
with the 4] neutron configuration.}
 \end{centering}
\end{figure*}
\begin{figure}
 \begin{centering}
\includegraphics[clip=true,width=0.48\textwidth,trim=0 0 0 0]{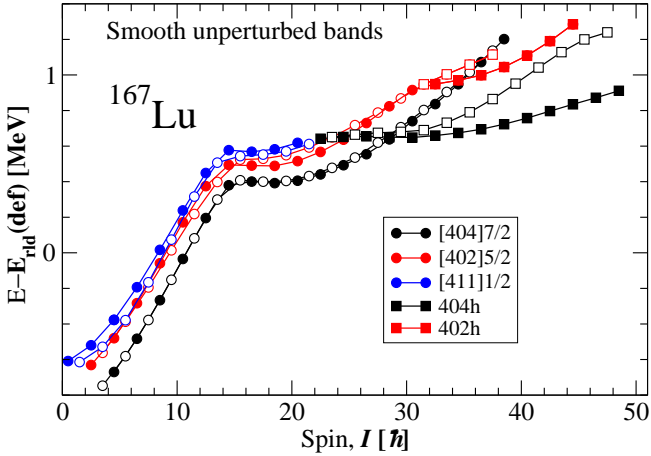}
\caption{\label{exp4} Experimental bands 
    built on the ${\cal N}=4$ proton band heads which will be compared with CNS and CNSB calculations.}
 \end{centering}
\end{figure}

 
In the previous section, the crossing between bands 1 and 10 at $I \approx 30$
was analyzed. It is now instructive to analyze the positive one-quasiparticle
bands in some more detail, i.e. the bands based on band heads with the odd proton
in a positive parity orbital. As seen from the single-routhian orbitals in
Fig.\ \ref{sp} there are three such orbitals close to the Fermi surface, the
[404]7/2 and [402]5/2 orbitals of $d_{5/2}g_{7/2}$ $(dg)$ origin and the 
[411] 1/2 orbital of $s_{1/2}d_{3/2}$ $(sd)$ origin. The experimental bands based on these
band heads are drawn in Fig.\ref{exp4}. As can be concluded e.g.\ from 
Fig.3 in Ref.\cite{rou15} it is evident which states should be
assigned to the different band heads at low spin. Furthermore, the lowest
band, starting on the [404]7/2 band head, remains separate from the other bands up to
the crossing at $I \approx 30$, i.e. it must be assigned as [404]7/2 in this
full spin range. 
In general, the [404]7/2 and [402]5/2 Nilsson orbitals have
very similar properties because they have both $n_z = 0$, i.e.\ they have
all quanta in the perpendicular direction. Furthermore, because they have
relatively high spin values along the symmetry axis, $\Omega = 7/2$ and $5/2$, 
respectively,
they will only give a small spin contribution as seen from the small slopes in
Fig.\ \ref{exp4}, where we can also see that the two signatures remain degenerate
up to the highest frequencies shown in the figure.  
Thus, the band which follows
the same trend as the [404]7/2 band but $\approx 200$ keV excited in energy must be
assigned as [402]5/2. 
This is in contradiction to the band assignment according to
Ref.\ \cite{rou15}. This difference is understood from the fact the
$\frac{11}{2}^-$ states of the [402]5/2 and [411]1/2 bands are almost degenerate and it is 
unclear how the bands should be connected. Compared with  Ref.\ \cite{rou15},
we have thus formed a band-crossing at $I=11/2$. The way we have formed bands is thus
mainly based on the energies. $M1$ transitions are observed both within the
bands and connecting the bands indicating that these bands are strongly mixed.
In any case,
for a general understanding of the bands within our cranking formalism,
the labeling in Fig.\ \ref{exp4} is certainly preferable.
This means that comparing with Fig.\ \ref{exp},
we can conclude that 
\begin{itemize}
\item Band [404]7/2 is composed of band 1 up to $I \approx 30$ and then by band 10. 
\item Band [402]5/2 in the spin range $I \approx 5.5 - 30$ is composed by bands 9a and 8b,
see Fig.\ \ref{exp}.
\item Band [411]1/2 in the spin range $I \approx 5.5 - 21.5$ is composed of 8a and 9b
\item The band which we label 404h (where h refers to `high spin') is composed
of band 10 in the spin range $I \approx 22.5 - 30$ and 
from band 1 for spin values above  $I \approx 30$.
\item Band 402h is built from band 9a and 8b above $I \approx 30$
\end{itemize}

\section{CNS and CNSB calculations}
\label{model}
\subsection{Formalism}

In the cranked Nilsson-Strutinsky (CNS) model \cite{beng85,Afa99,carl06} the Hamiltonian 
is taken as 
\begin{equation}
 H = H_{MO}(\varepsilon_2,\gamma,\varepsilon_4) - \omega j_{x},
\end{equation}
where $H_{MO}$ denotes the modified oscillator Hamiltonian \cite{Nil95} and 
$\omega j_x$ is the cranking term. The
renormalized total energies are calculated as the sum of the rotating
liquid-drop energy and the shell energy using the Strutinsky method \cite{strut}. The Lublin-Strasbourg
drop (LSD) model \cite{pomo03} is used for the static liquid drop energy while a radius parameter
$r_0 = 1.16$ fm and a diffuseness parameter $a=0.6$ fm is used for the rigid body moment of inertia \cite{carl06}.

In addition to the unpaired CNS calculations, we have also carried out
calculations in the cranked Nilsson-Strutinsky-Bogoliubov (CNSB) formalism with pairing included \cite{carl08,ma14}. The CNSB
model is based on the ultimate cranker formalism \cite{beng89,beng90}, using the same modified oscillator
potential as the CNS model, plus a monopole pairing term. The Hamiltonian for either protons or neutrons can be written as
\begin{equation}
 H = H_{MO}(\varepsilon_2,\gamma,\varepsilon_4) - \omega j_{x} + \Delta (P^{\dagger}+P) - \lambda \hat{N},
\end{equation}
where $P^{\dagger}$ ($P$) and $\hat{N}$ are the pair creation (annihilation) and particle number operators, 
respectively. In this
approach, the microscopic energy after particle number projection,
is minimized in a mesh of the pairing parameters, Fermi energy $\lambda_p$ and $\lambda_n$,
and pairing gap $\Delta_p$ and $\Delta_n$, as well as in the deformation space. As seen in Eqs.(1) and (2), 
the CNS and CNSB Hamiltonians are different only in the pairing terms.  Both
CNS and CNSB calculations are carried out in a quadrupole and
hexadecapole deformation mesh $(\varepsilon_2, \gamma, \varepsilon_4)$ 
with the same $\kappa$ and $\mu$ parameters.

For the $\kappa$ and $\mu$ Nilsson model parameters, we have used the
set StA150 \cite{rag15} which is the average of $A = 150$
\cite{beng90} and standard \cite{beng85} $\kappa$ and $\mu'$
parameters.  This is motivated by the fact that the $A = 150$
parameters have been fitted for nuclei with $N \approx 90$, and standard
parameters are more appropriate for the well deformed nuclei in the
middle of the rare-earth region, where the nucleus $^{167}$Lu with
$Z=71$ and $N=96$ is somewhere between these two regions.  
The StA150 parameters are listed in
Table.\ref{tab} for the shells where the two sets are not
identical, ${\cal N}=4$, 5 and 6.  
\begin{table}
\begin{center}
\caption{\label{tab} StA150 $\kappa$ and $\mu$ parameters for ${\cal N}=4$, 5, and
6 proton and neutron shells. These parameters are an arithmetic average of the
standard and $A=150$ parameters where it is only the values written in bold which
are different in the two sets.}
\begin{tabular}{ c c c c c }
\hline\hline
  & $\kappa_p$ & $\mu_p$ & $\kappa_n$ & $\mu_n$ \\ 
\hline 
 ${\cal N}=4$ & {\bf 0.0675} & {\bf 0.5337} & 0.07000 & 0.3900\\  
 
 ${\cal N}=5$ & 0.0600 & {\bf 0.6000} & 0.06200 & 0.4300 \\   
 
 ${\cal N}=6$ & 0.0540 & 0.6000 & 0.06200 & {\bf 0.3700}\\
\hline\hline

\end{tabular}
\end{center}
\end{table}

The only preserved quantum numbers in the CNSB formalism are parity $\pi$
and signature $\alpha$ for protons and neutrons, i.e.\ the configurations
can be labeled as $(\pi_{p},\alpha_p)(\pi_{n},\alpha_n)$. The advantage of
the CNS calculations is mainly that configurations can be fixed in a much
more detailed way. Thus, the diagonalization is carried out in a 
rotating harmonic oscillator basis \cite{beng85} where the small couplings 
between the ${\cal N}_{rot}$-shells are neglected making ${\cal N}_{rot}$ a preserved
quantum number. Furthermore, within the ${\cal N}_{rot}$-shells, it is possible
to define approximate quantum numbers where it is generally straightforward
to distinguish between orbitals with their major amplitude, either in the intruder
high-$j$ shell or in the other `low-$j$' shells. Indeed, sometimes a further
distinction specifying orbitals as belonging to pseudo-spin partner orbitals
is also possible, see Fig.\ \ref{sp}(a). Here, we will mainly make a distinction between high-$j$
and low-$j$ which means that configurations in $^{167}$Lu can be labeled
as $[p_1(p_2 p_3);(n_1 n_2)n_3]$, where $p_1$ represents the number of
$h_{11/2}$ protons and $n_3$ the number of $i_{13/2}$ 
neutrons. The labels $p_2$ and $p_3$,
represent the number of protons of $f_{7/2} h_{9/2} $ and $i_{13/2}$
character while $n_1$ and $n_2$ are the number of neutron holes
in ${\cal N}=4$ and $h_{11/2}$ orbitals, respectively. These numbers 
in parentheses are specified only if nonzero. 
For $^{167}$Lu with 71 protons and 96 neutrons, configurations can be
written in full as
\begin{center}
$\pi{({\cal N}=4)^{-(p_1+p_2+p_3-1)}(h_{11/2})^{p_1} (f_{7/2} h_{9/2})^ {p_2}(i_{13/2})^{p_3}}$ \\
$\nu{({\cal N}=4)^{-n_1}(h_{11/2})^{-n_2}(f_{7/2} h_{9/2})^{14+n_1+n_2-n_3}(i_{13/2})^ {n_3}}$.
\end{center}
relative to a $Z=70$ proton core with all shells up to ${\cal N}=4$ filled and 
a  neutron core with all $j$-shells up to neutron number $N=82$ filled. 

The transitional quadrupole moment $Q_t$ is calculated from the 
deformation of a state, where the formula is given e.g.
in Refs. \cite{Juo00,Wan11}.

\subsection{Comparison of CNS and CNSB calculations}
\label{comp}
\begin{figure}
 \centering 
 \includegraphics[clip=true,scale=0.6,trim=0 0 0 0]{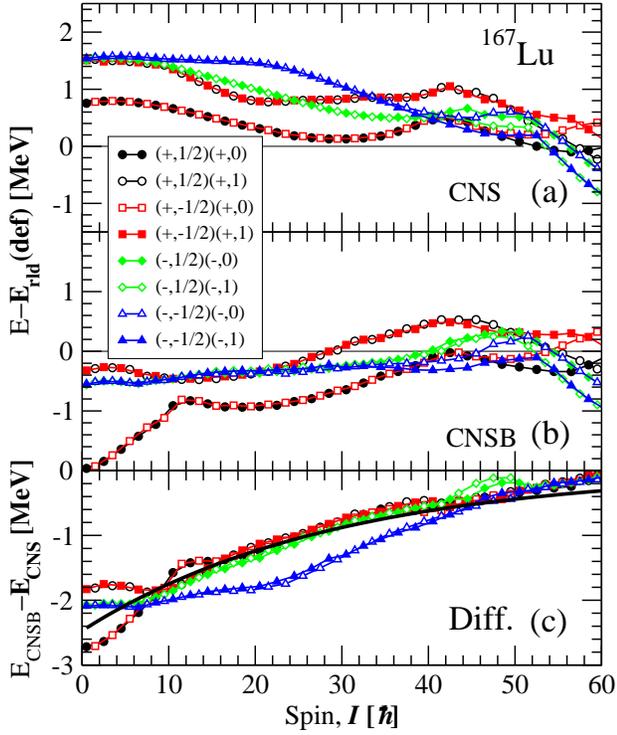}
\caption{Calculated unpaired (a) and paired (b) yrast bands and their differences
  (c) in $^{167}$Lu for the positive parity
  configurations. The average pairing energy is shown by a black line
in panel (c).}
\label{pos}
\end{figure}

\begin{figure}
 \centering 
\includegraphics[clip=true,scale=0.6,trim=0 0 0 0]{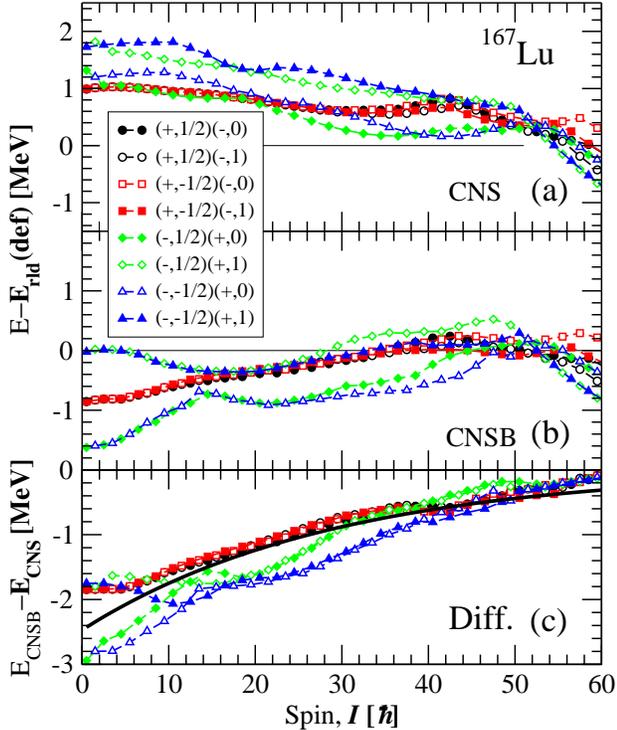}
\caption{Same as Fig.\ \ref{pos} but for  negative parity configurations.
}
\label{neg}
\end{figure}

 The comparison of the 16 CNS and CNSB yrast combinations designated with
 $(\pi_{p},\alpha_p)(\pi_{n},\alpha_n)$  is
 shown in the full spin range, $I=0-60$, in Figs.\ \ref{pos} and \ref{neg} for the positive and negative
 parity configurations, respectively.

The energy differences between the unpaired and paired calculations,
which correspond to the pairing correction energies for the
yrast configurations, are displayed
in Fig.\ \ref{pos}(c) and \ref{neg}(c). The band crossing 
(backbending) in the
paired CNSB yrast bands with the $(+,0)$ neutron configuration
is evident. It is well-known that the 
rotational alignment of one $i_{13/2}$ neutron pair is responsible for
this band crossing, which is usually seen in the rotational bands in the
rare-earth nuclei in the spin region $I=12-16$ \cite{jen81}. 


One can also see a clear trend for the pairing correction
energies in different configurations of $^{167}$Lu.  At higher spins
than $I \approx 20$, there are no significant discontinuities in the
differences between the CNSB and CNS yrast lines in $^{167}$Lu.  It
suggests that the calculated crossings at higher spin than $I \approx 20$ are
similar in two formalisms.  It indicates that these band crossings can be
understood as caused by crossings between unpaired CNS configurations. 

 In Figs.\ \ref{pos} and \ref{neg}, the configurations
with negative parity for the protons have a stronger pairing 
in the $I \approx 10-40$ spin range than
the other configurations. This is especially true for the 
$(\pi,\alpha)_{p} = (-,-1/2)$ configurations but to some extent also
for $(\pi,\alpha)_{p} = (-,1/2)$.
It appears that the reason
is that the orbital which is blocked in these configurations is far away
from the Fermi surface, see Fig.\ \ref{sp}(a).
Thus, at low frequencies, the [523]7/2 orbital is far below the Fermi
surface while the [514]9/2 orbital is far above. 
At the deformation used in Fig.\ \ref{sp}, it is only for somewhat higher
frequencies that the $\alpha = 1/2$, [541]1/2 orbital will come in the Fermi
energy region. The blocking of an orbital close to the Fermi energy will
reduce the pairing correlations while blocking of an orbital further
away will have a smaller effect. Thus, the calculated pairing
energies in the different configurations are consistent with
expectations based on the position of the orbitals around the Fermi
energy.

To get a more realistic results from the CNS calculations, we 
fit a smooth
function to the curves in Figs.\ \ref{pos}(c) 
and
\ref{neg}(c), corresponding to the average pairing correction energy.
Then it is straightforward to add this average pairing correction to
the CNS results.
A least square fit of an exponential function in the $I=0-60$ region of
spin results in
\begin{center}
 $E_{pair}=-2.47 \exp(-I/29).$
\end{center}
In the calculations presented below, this average pairing energy will
be added to the CNS energies.

\subsection{Expected low-lying configurations}

The routhians for protons and neutrons
presented in Fig.\ \ref{sp} can be used to get a general understanding of 
which configurations are expected to be observed in 
$^{167}$Lu. The proton diagram suggests that at low spin 
for positive parity, the proton valence space
configurations will have the odd proton in the
[404]7/2 or [402]5/2 orbitals which are pseudospin partners in the 
$d_{5/2}g_{7/2}$ ($dg$) shells. Also the [411]1/2 orbital of 
$s_{1/2}d_{3/2}$ ($sd$) origin comes rather close to the Fermi energy. 
For negative parity, the $h_{11/2}$ orbitals [523]7/2 and [514]9/2 are found
at a similar distance from the Fermi energy. It is clear that also the
[541]1/2 orbital of  $h_{9/2}f_{7/2}$ ($hf$) origin could be competitive.
For higher frequency, this $hf$ orbital and also the
$i_{13/2}$ [660]1/2 orbital comes close to the Fermi energy, where the
down-slope of the routhians shows that these orbitals will give a 
significant spin contribution.

Because of pairing,
the favored low-spin band heads will be formed when the proton valence space 
configurations are combined with an even parity neutron configuration.
The resulting {\em one-quasiparticle configurations} will
be dominated by $\nu(i_{13/2})^4$ as can 
be concluded from  Fig.\ \ref{sp}(b). However, this figure suggests that
also the  $(i_{13/2})^5$ configurations with  an odd neutron in the $(i_{13/2})$ 
as well as in the $hf$ orbitals will come low in energy. When they are combined with
the proton valence space configurations, {\it three-quasiparticle
configurations} are formed. They will become competitive in energy
around $I=10-15$, because after the first backbend, also
the $\nu(i_{13/2})^4$ configurations will have three `odd particles', i.e. they
will have a similar pairing energy as the three-quasiparticle configurations. 
It is then our experience that for higher spin values, 
the spin is mainly built from the gradual alignment of the valence particles
and it is not meaningful to describe them as having a fixed number of quasiparticles.
     

The main features of the level scheme can be understood from the discussion above.
However, it is also possible to read out some additional features from Fig.\ \ref{sp}.
Thus, the neutron [505]11/2 orbital of $h_{11/2}$
origin comes close to the Fermi surface, which means that configurations with one or
two holes in this orbital might become competitive. It appears that no
configurations with one hole is observed in $^{167}$Lu, while for
configurations with two holes, it is
clear that 6  $i_{13/2}$ orbitals will be occupied, see Fig.\ \ref{sp}(b). 
Note, however, that the sixth $i_{13/2}$ orbital with 
signature $\alpha = -1/2$ goes away from
the Fermi energy with increasing spin. Thus, it is only for low spin values
that  $\nu(h_{11/2})^{-2}(i_{13/2})^6$ configurations are competitive in energy.
Furthermore, with no $h_{11/2}$ holes,  
for $(i_{13/2})^5$ configurations 
it is only those with
the 5th neutron in the signature  $\alpha = 1/2$ branch which are competitive
in energy, while $(i_{13/2})^6$ 
configurations will come too high in energy to be of 
practical interest. 
The conclusion is thus that there are only four neutron configurations of interest,
namely  $(i_{13/2})^4$, $(i_{13/2})^5$ with two signatures for the
$hf$ neutron and $(h_{11/2})^{-2}(i_{13/2})^6$. 
Note the signature degeneracy for the [523]5/2 orbital in
Fig.\ \ref{sp}(b), i.e.\ signature degenerate bands are expected in the 
$(i_{13/2})^5$ configurations.

\section{Interpretation of the bands}
\label{analyze}
In our analysis of the observed bands, we will first consider those which start
out as one-quasiparticle bands, namely those which are built on ${\cal N} = 4$
orbitals in Sec.\ \ref{n4}, those built on $h_{11/2}$ orbitals in Sec.\ \ref{h11}
and those based on the [541]1/2 orbital of $hf$ origin in Sec.\ \ref{541}. For the
latter band, which is yrast at low spin, we will conclude that up to $I \approx 20$,
it is based on a configuration with two $h_{11/2}$ neutron holes. Then in Sec.\ \ref{sect-3qp}, 
we will consider the bands of both parities which are assigned to
start out as three-quasiparticle bands. We may note that one of the 
three-quasiparticle bands is only seen at high spin and because it 
feeds into the ${\cal N} = 4$ bands, it will be
treated in Sec.\ \ref{n4}.

\subsection{Bands based on the ${\cal N}=4$ band heads}
\label{n4}

\begin{figure}
 \begin{centering}
\includegraphics[clip=true,width=0.48\textwidth,trim=0 0 0 0]{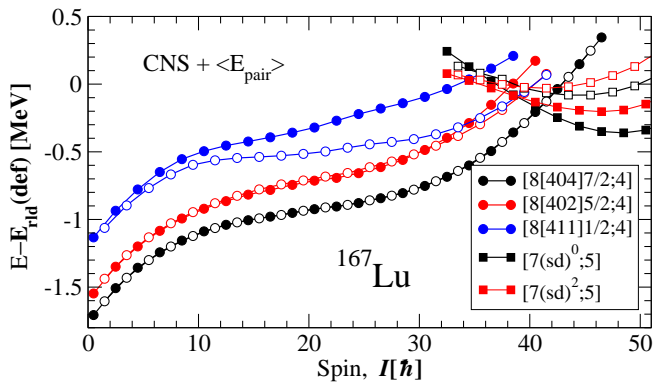} 
  \caption{\label{calc4} CNS configurations which are
    assigned to the experimental bands in Fig.\ \ref{exp4}.}
 \end{centering}
\end{figure}

The observed bands based on the proton ${\cal N}=4$ orbitals are shown in Fig.\ \ref{exp4}.
The configurations calculated in CNS which are assigned to these bands are drawn
in Fig.\ \ref{calc4}. 
The observed and calculated bands show the same structure with the 
[404]7/2 band as yrast and then [402]5/2 and [411]1/2 as the first 
and second excited bands. The observed bands for spins up to
$I \approx 30$ are only spread within 300 keV while the calculated
ones are somewhat more spread, i.e.\ within 600-700 keV. This is however
a rather small difference and it would only require a small change
of the single-particle parameters to put the [411]1/2 orbital closer to
the Fermi energy, see Fig.\ \ref{sp}(a). All the observed bands go through a backbend
at $I \approx 14$ which is of course the AB crossing where two $i_{13/2}$
neutrons align. Then, at $I \approx 30$, band crossings are
observed both in the [404]7/2 and [402]5/2 bands. In Ref.\ \cite{rou15}, it
is suggested that this crossing in the [404]7/2 band is caused by another
alignment among the $i_{13/2}$ neutrons, the CD crossing. The idea is thus
that the lowest energy band, labeled 404h, has four $i_{13/2}$ neutrons 
aligned while the
upper band has only 2 such neutrons aligned. Our experience is that, if 
such configurations could be formed, the band
with only two neutrons aligned would be very unfavored in energy and it
seems very strange that this band can be followed for approximately 10
units of angular momentum up to $I \approx 40$, where  
this higher band is only about 200 and 400 keV above the two signatures
of the bands with
four $i_{13/2}$ neutrons aligned. Even more strange is, however, that the
yrast band shows an appreciable signature splitting while the excited
band does not. Because in the full spin range, the odd particle will be
in the signature degenerate [404]7/2 orbital, see Fig.\ \ref{sp}(a), no such
splitting is expected as also noted in Ref.\ \cite{rou15}. Furthermore,
the calculations including pairing which are presented below do not show
any more crossing after the AB crossing. Our interpretation is then that
after the first AB crossing, the pairing energy is severely quenched so
that no CD crossing is observed; instead the remaining alignment among
the $i_{13/2}$ neutrons occurs gradually as described by the unpaired
CNS calculations. 


\subsubsection{Interpretation of the band crossing at $I \approx 30$}
\label{n4-cross}

The question is then how the band labeled 404h which intersects the [404]7/2 band
at $I \approx 30$ should be understood. From the single-routhian 
diagrams in Fig.\ \ref{sp} one could expect a band where one [523]7/2
proton is lifted to the [541]1/2 orbital. The corresponding band 
labeled [7(10);4], not shown in Fig.\ \ref{calc4},
will cross the lowest [8;4] band
somewhat above $I=40$. However, this band will have the odd proton in the [404]7/2
orbital which means that two signature degenerate bands would be expected
contrary to experiment. Indeed, another band with negative parity for
both protons and neutrons, $\pi(h_{11/2})^7\nu (i_{13/2})^5$ or [7;5] will cross 
the lowest [8;4] band, the [404]7/2 band, at a slightly lower spin value around
$I=40$, see Fig.\ \ref{calc4}. 
Note that with this interpretation for the 404h band, its signature splitting
is about the same as in the calculated band assigned
to it, the [7;5] band. The signature
splitting is caused by the odd $hf$ neutron with the odd $h_{11/2}$ proton
in its favored signature. An alternative would be to
leave the $hf$ neutron in the favored signature and form the two signatures
from the odd $h_{11/2}$ orbital instead. However, the unfavored
signature branch would then come at a higher energy and the signature splitting
would be considerably larger in calculations than in experiment, see Fig.\ \ref{3qp} below.  

It is seen in Fig.\ \ref{exp4} that there is a similar
crossing in band [402]5/2 as in band [404]7/2, also 
at spin $I \approx 30$ which suggests a similar
structure change in the two bands. Indeed, as becomes evident
from Fig.\ \ref{sp}(a), with 7 protons in the
$h_{11/2}$ shell, the ${\cal N}=4$ protons can have different distributions
over 
the $sd$ and $dg$ orbitals.
Note also that the calculated deformation at $I \approx 40$ 
for the [7;5] configuration is
$\varepsilon_2 \approx 0.22$ where the  $sd$ orbital, [411]1/2, and the
$dg$ orbitals, [404]7/2 and [402]5/2, come much closer in energy than at
the deformation in  Fig.\ \ref{sp}, $\varepsilon_2 = 0.27$.

It turns out that with constraints on the distribution of protons in the
$dg$ and $sd$ orbitals, respectively, those with a closed $Z=64$ core,
$(sd)^0$ and those with two $(dg)$ holes in the core, i.e.\ two $(sd)$ protons,
$(sd)^2$, have a very similar energy, see Fig.\ \ref{calc4}. However, these two
configurations have a different deformation, $\gamma \approx -30^{\circ}$ and 
$\gamma \approx -10^{\circ}$, which means that the division into 
$(sd)$ and $(dg)$ orbitals is not so well-defined. 
In general, such a division appears to be pretty straightforward for
positive values of $\gamma$ and $\gamma \approx 0$ but more arbitrary for large
negative values of $\gamma$ where the two highest $dg$ orbitals and
the lowest $sd$ orbital are more strongly
mixed. Thus, the energy difference between the  $(sd)^0$ and $(sd)^2$ bands in
 Fig.\ref{calc4} is uncertain but these calculations show that there is room
for two [7;5] bands to explain the observed 404h and 402h bands. 

A problem is that the 
crossings are observed at $I \approx 30$ while, they are calculated at
$I \approx 40$. However, as noted below, with pairing included the 
calculated crossing will come at a lower spin value closer to 
experiment.

\subsubsection{CNSB calculations}
\label{cnsb}
\begin{figure}
 \begin{centering}
\includegraphics[clip=true,width=0.45\textwidth,trim=0 0 0 0]{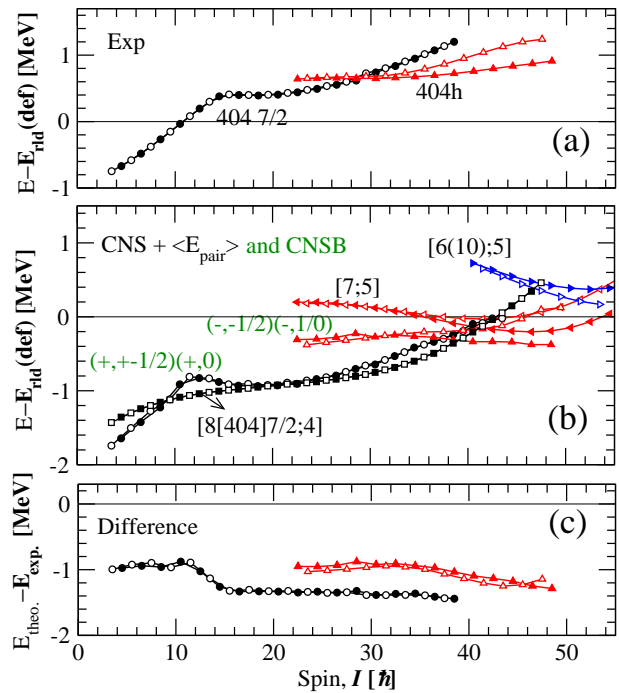} 
  \caption{\label{expth-404} Experimental energies (a) and theoretical 
`CNS + average pairing' and CNSB energies (b) relative to a rotating liquid drop and the differences between the experiment and the 
CNSB yrast configurations (c) as a function of spin
 for ${\cal N}=4$ configurations.}
 \end{centering}
\end{figure}
It is only for the bands which are yrast in the groups with 
parity and signature fixed for protons and neutrons that it is
straightforward to compare with paired calculations. Thus, for the
 bands which are lowest in energy in Figs.\ \ref{exp4} and \ref{calc4}, 
such a comparison is provided in Fig.\ \ref{expth-404}.
In this figure, the observed bands are shown relative to the rotating
liquid drop energy in the upper panel, the corresponding calculated
bands in CNS with average pairing and CNSB are shown
in the middle panel with the difference between experiment and CNSB
calculations in the lower panel. Comparing the two calculations in the
middle panel, it turns out that the CNSB pairing will be stronger than
the average for the [7;5] bands while, for $I=30-40$, it will be weaker than the average
in the [8;4] bands. Consequently, relative to CNS with average pairing,
the crossing between the two bands will come at a lower spin in CNSB and
thus closer to experiment. Furthermore, the two difference curves in the
lower panel comes relatively close together, separated by approximately
400 keV which is clearly within the general uncertainty of the calculations.
Thus, from relative energies and also from the signature splitting, the 
description of this crossing at the yrast line is well described as a
crossing between an [8;4] and a [7;5] configuration. Concerning the
interaction strength between these two rather different configurations,
it is somewhat more difficult to judge if the observed value around 30 keV,
see Sec.\ \ref{undis}, is what would be expected. 


\subsubsection{Alignments}

\begin{figure}
 \begin{center}
\includegraphics[clip=true,width=0.4\textwidth,trim=0 0 0 0]{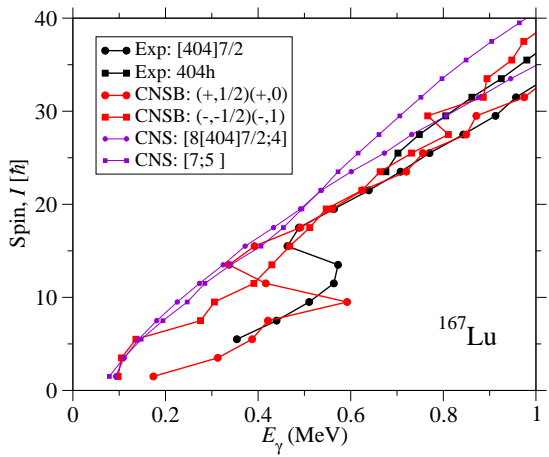}
  \caption{\label{align4} Spin $I$ versus transition energy $E_{\gamma}$ for the
$\alpha = 1/2$ branches of the observed bands [404]7/2 and 404h are 
compared with calculations for the configurations assigned to these bands 
in the CNS and CNSB formalisms.  
}
 \end{center}
\end{figure}

The alignment of the observed $\alpha = 1/2$ branches of bands 1 and 10 is discussed in the appendix.
When we let these bands cross at $I \approx 30$, they are referred to as
bands [404]7/2 and 404h, see Fig.\ \ref{exp4}. Their spin values are
drawn versus the transition energy in Fig.\ \ref{align4}.  
The first AB crossing in the [404]7/2 band has been studied in detail
previously, see e.g. \cite{Yu90}. It is however satisfying to observe
that it is well explained in the present CNSB calculations while this
paired crossing is of course not reproduced in the CNS calculations.
Our main interest here is however the spin difference between the 
[404]7/2 and the 404h bands when they cross around $I \approx 30$,
where the observed value is around $2.5 \hbar$ as determined in the
appendix and which can also be read out from Fig.\ \ref{align4}. It
is then very satisfying, that the CNSB alignments reproduce the 
experimental values for both bands in an extended frequency range, 
and thus also their difference,
on the average. The rather large fluctuations in the calculated 
values can probably be explained by the fact that we do not interpolate
in the pairing space, i.e. the energy is taken as the lowest value in
the $(\Delta, \lambda)$ mesh.  

The values calculated in the CNS formalism are also drawn in Fig.\ \ref{align4}.
It is interesting that also at high spin, $I=30-40 \hbar$, the values
calculated in CNS and CNSB are rather different. If an average pairing
is added, the CNS results will of course come closer to CNSB (and experiment).
However, for example the spin difference between the [8;4] and [7;5]
configurations is still clearly larger than for the corresponding CNSB
configurations and thus also more different from experiment. This appears
consistent with the observation from Fig.\ \ref{expth-404} that the different
pairing for the  [8;4] and [7;5] configurations is important to get a
satisfactory agreement between experiment and calculations.

\subsubsection{Higher spin discontinuities}
In the unfavored signature of the observed 404h band at the highest spin,
a small disturbance is seen suggesting an interaction or crossing with some
other bands. Indeed different bands come down at high spin, for example with
one proton excited to [541]1/2 as shown in Fig.\ \ref{expth-404}(b) 
and at a similar energy,
bands with a proton excited to the [660]1/2 orbital (not shown in
Fig.\ \ref{expth-404}(b)). We may note,
however that the crossing between the $\alpha = -1/2$ [7;5] and [6(10];5] 
bands in Fig.\ \ref{expth-404}(b) does not only correspond to lifting a proton 
from the [523]7/2 to the [541]1/2 orbital but also to a signature change not
only for  this proton but also for the negative parity neutron. Furthermore, 
while it is difficult to identify any disturbance of the $\alpha = 1/2$ 
[7;5] band from Fig.\ \ref{expth-404} such a disturbance is seen in the 
relative alignment plots in Fig.\ \ref{align-b12}, i.e.\ this plot suggests
a similar disturbance in both signatures of the [7;5] band. 
Indeed, a closer
look on the [7;5] bands reveals a shape change from $\gamma \approx 0$ to
$\gamma$ approaching $-30^{\circ}$ in both signatures of the [7;5] band.
This shape change is clearly related to the different shapes of the two
[7;5] bands, $(sd)^0$ and $(sd)^2$, discussed above.
The effect on the energies from this shape change is seen rather clearly
at $I \approx 46$ in the $\alpha = -1/2$ but not in the $\alpha = 1/2$
branch of the calculated [7;5] band in Fig.\ \ref{expth-404}. Thus, our
interpretation is that this shape change causes the
small discontinuity observed at $I \approx 46$ in the 404h band
drawn in Fig.\ \ref{expth-404}(a).

\subsubsection{Bands 6 and 7}
Band 7 decays to the [404]7/2 band with two rather intense
transitions, $14.5 \rightarrow 12.5$ and $16.5 \rightarrow 14.5$.
In addition, a weak transition is observed to band 4, $26.5
\rightarrow 24.5$, which can be understood because especially
the $I=24.5$ states of bands 7 and 4 come close together. The
transitions to the [404]7/2 band suggests some relation with this
band and the energy curve of band 7 suggests that it can be understood as a
continuation of the  [404]7/2 band with no AB alignment. 
This might suggest that the crossing seen in  band 7 at $\hbar\omega \approx 0.32$ MeV can be understood as a BC alignment. The energy curve of band
6 and its decay to the [402]5/2 $\alpha = -1/2$ band suggests a
similar scenario for this band, namely that it could be the $\alpha = -1/2$ 
continuation of the [402]5/2 band with no AB alignment.  However, the
fact that no BC crossing is seen and the fact that band 6 has several
links to a negative parity 541 band makes the interpretation very
tentative.


\subsection{The $h_{11/2}$ band (band 15)}
\label{h11}
\begin{figure}
 \begin{center}
\includegraphics[clip=true,width=0.4\textwidth,trim=0 0 0 0]{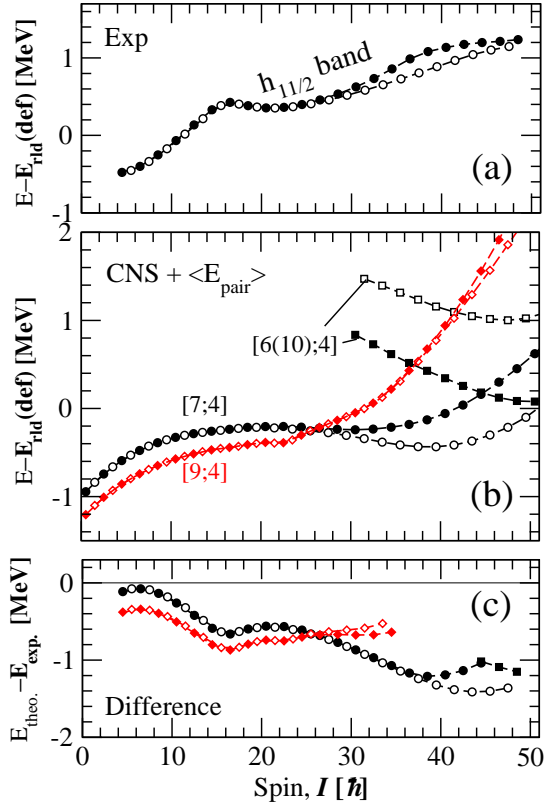}
  \caption{\label{expth15} Experimental energies (a) and theoretical 
`CNS + average pairing' energies (b) relative to a rotating liquid drop and their differences 
(c) as a function of spin
 for the $h_{11/2}$ band in $^{167}$Lu.}
 \end{center}
\end{figure} 
The signature partner $h_{11/2}$ bands (Bands 15 in Ref.\ \cite{rou15})  
are the lowest negative-parity bands in
$^{167}$Lu for spin values above $I \approx 18$. 
The CNS configurations with either 7 or 9 $h_{11/2}$ protons,
[7;4] or [9;4] and thus  with the odd proton in either 
[523]7/2 or [514]9/2 come at a very similar energy at low and
intermediate spin, see Fig.\ \ref{expth15}. Thus, 
at low spin, it is somewhat uncertain if the $h_{11/2}$ band should
be assigned to one or the other of these two bands or maybe rather
a mixture of them.
At higher spin values, the [7;4] configuration
is clearly favored in energy and also the
signature splitting is consistent with the [7;4] interpretation. Thus,
because no discontinuity is seen in the observed band in the 
$I \approx 16-40$ spin range, it appears well established that above
the backbend at $I \approx 16$, where pairing is severely quenched, the
$h_{11/2}$ band should be assigned to the [7;4] configuration, i.e.\ with the
odd proton in the  [523]7/2 orbital. This is contrary to Ref.\ \cite{rou15}, 
where the $h_{11/2}$ band is assigned to a
configuration with the odd proton in the [514]9/2 orbital.  

Our
calculations suggest that the observed crossing in the 
$\alpha = 1/2$,  $h_{11/2}$ band at $I \approx 40$ is caused by
the excitation of a proton from [523]7/2  $h_{11/2}$ 
orbital to the [541]1/2 $h_{9/2}$ 
orbital at the frequency $\hbar \omega = 0.5$ MeV 
where these orbitals cross,  
see Fig.\ \ref{sp}(a). The observed crossing in the $h_{11/2}$ band
is thus explained as a crossing between the $\alpha = 1/2$
[7;4] and [6(10);4] configurations in the CNS labeling.
A similar crossing is observed in $^{168}$Hf \cite{yad09,kar12}. 
Our interpretation is in agreement with Ref.\ \cite{rou15} where this
crossing is labeled as an $fg$ crossing.
Note
that no corresponding crossing is expected for signature 
$\alpha = -1/2$. This is understood from the high energy
of the $\alpha = -1/2$ [541]1/2 orbital in Fig.\ \ref{sp}(a)
corresponding to a high energy of the corresponding 
[6(10);4] band, see middle panel of Fig.\ \ref{expth15}.

\begin{figure}
 \begin{centering}
\includegraphics[clip=true,width=0.4\textwidth,trim=0 0 0 0]{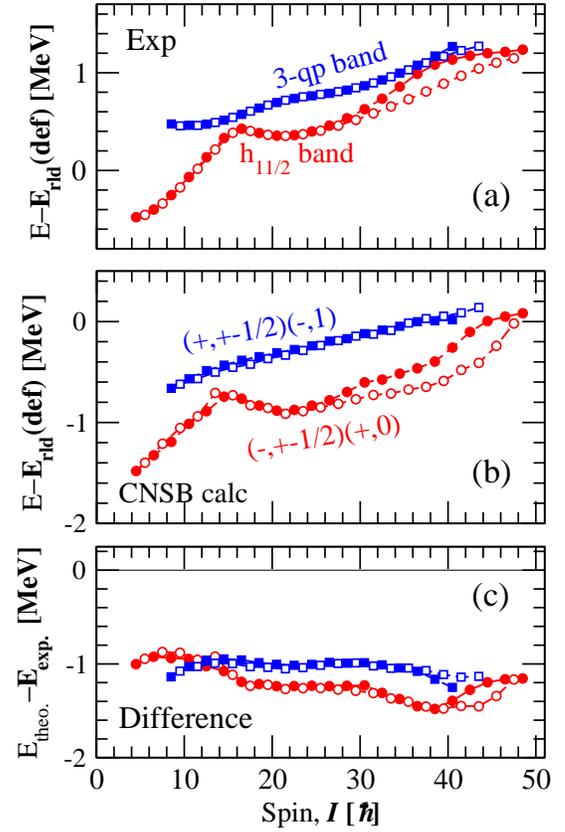} 
  \caption{\label{expth2} Experimental energies (a) and theoretical 
CNSB energies (b) relative to a rotating liquid drop and their differences (c) as a function of spin
 for the $h_{11/2}$ band (Sec.\ \ref{h11}) and the three-quasiparticle band 2 
(Sec.\ \ref{sect-3qp}) in $^{167}$Lu.}
 \end{centering}
\end{figure}

Considering the one-quasiparticle configurations with negative parity,
$(-,\pm 1/2)(+,0)$, those with the odd proton in the [541]1/2
orbital of $hf$ origin are generally lower 
in energy than those with an odd $h_{11/2}$ proton in the CNS
calculations. However, as noted in Sec.\ \ref{comp}, pairing is stronger in the
`$h_{11/2}$ configurations'. Therefore in the CNSB calculations, the 
$(-,\pm 1/2)(+,0)$ yrast configurations has 
in general the odd particle in $h_{11/2}$. Thus, the observed 
$h_{11/2}$ band can be compared with the full
pairing CNSB configurations as done in Fig.\ \ref{expth2}. 
The difference between experiment and calculations is now 
more constant at a value around or just below $-1$ MeV.
As would be expected,
it is especially below the first $i_{13/2}$ band-crossing where
the average pairing energy comes out too small.

When checking the CNSB calculations in
more detail, it turns out that for signature $\alpha = 1/2$,
the [541]1/2 band might come
slightly below the $h_{11/2}$ band for some spin values.
However, because the two bands have a very similar
energy in  these cases, the error which is introduced because
the odd proton is placed in the `wrong orbital' is so small that
it is not noticeable in Fig.\ \ref{expth2}.  
Note also that for signature $\alpha = -1/2$,
the $h_{11/2}$ band is clearly calculated lowest in energy.
Therefore, the fact that the two signatures are degenerate
in an extended spin range in Fig.\ \ref{expth2} shows that
also the branch with signature $\alpha = 1/2$ is in general
built with an odd $h_{11/2}$ proton.  
The competition between the $hf$ and $h_{11/2}$ one-quasiparticle 
bands will be further discussed in Sec.\ \ref{sqt} below.

\subsection{The 541 bands (bands 13,14)}
\label{541}
\subsubsection{The lowest 541 band in $^{167}$Lu and neighboring nuclei}
\begin{figure}
 \begin{center}
\includegraphics[clip=true,width=0.42\textwidth,trim=0 0 0 0]{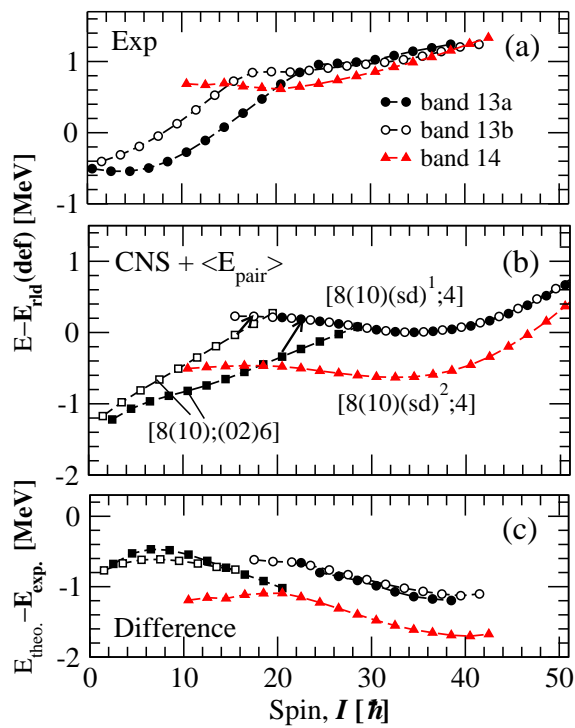}
  \caption{\label{expth1314} Experimental energies (a) and theoretical 
`CNS + average pairing' energies (b) relative to a rotating liquid drop and their 
differences (c) as a function of spin
 for the 541 bands in $^{167}$Lu, i.e. the  bands 13a, 13b and 14.}
 \end{center}
\end{figure}

The other low-lying one-quasiparticle negative parity configuration is 
built with the odd proton in the [541]1/2 orbital. It turns out
that with this proton configuration, the neutron configuration with
two holes in $h_{11/2}$ orbitals is favored at low spin, see Fig.\ \ref{expth1314}(b). Furthermore, the low spin ranges of 
the 13a,b bands are well described by this $\nu(h_{11/2})^{-2}$
configuration. In the calculations, the favored $\alpha = 1/2$
signature is crossed by the corresponding configuration with no
neutron holes at $I=18$. Comparing with experiment, a very similar
crossing is observed between the bands 13a and 14,
i.e.\ this crossing is well
understood in the unpaired formalism. Note, however, that also in
the unpaired formalism, this corresponds to a neutron $i_{13/2}$ crossing;
namely a configuration change from $\nu(i_{13/2})^{6}$ to  
$\nu(i_{13/2})^{4}$ and that the $i_{13/2}$ neutrons are expected to
contribute with considerably more spin in the latter than in the
former configuration. In this way, this unpaired crossing has 
much in common with a paired $i_{13/2}$ crossing. In some sense,
this is analogous to the description of the $h_{11/2}$ proton
crossing in $N \approx 90$ nuclei as a crossing between configurations
differing by two protons in the  $h_{11/2}$ orbitals, see e.g.\ 
Ref.\ \cite{Ben85}

It is interesting that the crossing bands 13a and 14 can
be followed both before and after the crossing which is another
indication that this two-level crossing can be treated in the unpaired formalism. 
Indeed, with moments of 
inertia which vary linear with spin $I$ and with a coupling strength
of 24 keV, the observed bands are fitted within $\pm 3$ keV by
two interacting smooth bands.

\begin{figure}
 \begin{centering}
\includegraphics[clip=true,width=0.48\textwidth,trim=0 0 0 0]{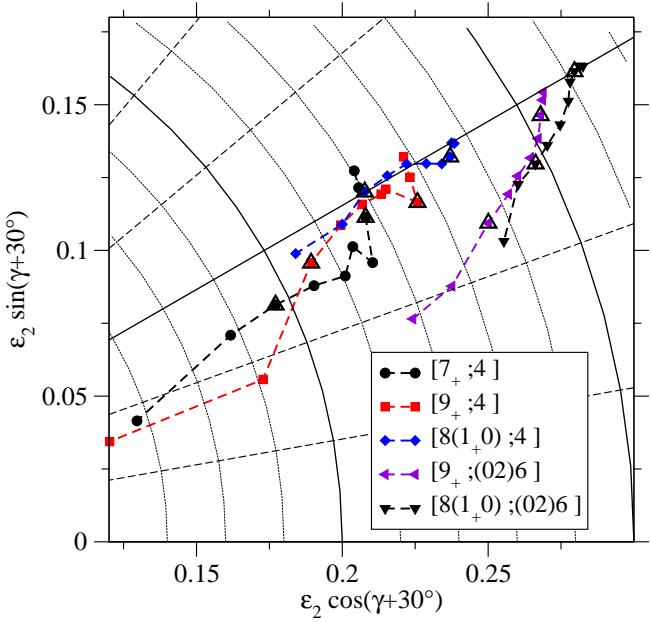} 
  \caption{\label{defh9} Deformation trajectories calculated
in the CNS formalism for some negative parity 
configurations in $^{167}$Lu, drawn to illustrate the polarization effects of
a [541]1/2 proton and two $h_{11/2}$ neutron holes. The trajectories are drawn in
steps of $6 \hbar$ in the spin range $I=4.5-58.5$. The deformation within these
fixed configurations will decrease with increasing spin where the points for
$I=16.5$ and 46.5 are highlighted by big triangles. In the legend, the subscript
`+' specifies the signature for the odd proton, $\alpha = 1/2$.
}
 \end{centering}
\end{figure}

Rotational structures built on the proton configuration 8(10),
i.e.\ with the odd proton in the [541]1/2 orbital, are
observed systematically in the neighboring nuclei around $^{167}$Lu,
for example in the
isotopes $^{165}$Lu \cite{sch04} and $^{169}$Lu \cite{oga93} or the
isotone $^{169}$Ta \cite{hart06}.
The properties of such bands were discussed in Ref.
\cite{Jen01}, where the nucleus $^{165}$Tm was
studied in detail. In general, the backbending in [541]1/2 bands is 
delayed which is partly understood as caused by an increased
deformation of these bands. However, the polarization effect of
the [541]1/2 orbital appears to be too small as seen for example
from measured transition quadrupole moments in the $^{165}$Tm 
nucleus \cite{Jen01}. With two holes in the neutron $h_{11/2}$
orbitals the polarization effects will be much larger as seen
from the calculated deformations presented in Fig.\ \ref{defh9}.
The figure shows that for one-quasiparticle configurations with the odd
particle in $h_{11/2}$, the calculated quadrupole deformation 
at low spin values is
$\varepsilon = 0.24-0.26$ with an increase to 
$\varepsilon = 0.27-0.28$ with the odd proton in the [541]1/2
orbital and then a larger increase to 
$\varepsilon = 0.30-0.32$ with two $h_{11/2}$ neutron
holes. This increased deformation might help to understand
the larger crossing frequencies in the [541]1/2 bands in the
$A=160-170$ mass region. Note however that for $^{167}$Lu,
the observed crossing is rather described as caused by a
crossing between `unpaired orbitals'. 

\subsubsection{The 541 bands at higher energy}

It remains to interpret the band crossings in the 13a,b bands
at spin values $I \approx 22$ and $I \approx 15$, where the crossing
frequencies can be read out in Fig.\ 17 of Ref.\ \cite{rou15} as 
$\hbar \omega \approx 0.38$ and  $\hbar \omega \approx 0.28$ MeV, respectively.
The first idea would be that this is a standard AB crossing within the
[8(10);(02)6] configuration
but this appears very unlikely because then the two 
signatures should cross at the same frequency. Furthermore, the fact
that the two signatures are essentially degenerate in energy after the
crossing excludes the assumption that they are built on the two signatures
of the [541]1/2 orbital because a large signature splitting is calculated for
this orbital, see Fig.\ \ref{sp}(a). Indeed, the fact that these two bands are
almost as low in energy as band 14 at $I \approx 40$ makes it very
unlikely that they are built with two $h_{11/2}$ neutron holes above
the crossing, because such configurations are calculated to come high in
energy for high spin values. Similarly, the low energy 
makes it very unlikely that they are built on the unfavored
signature of the [541]1/2 orbital. Instead, it appears that they are both 
built on the favored signature of the [541]1/2 orbital. Indeed considering
the proton orbitals in Fig.\ \ref{sp}(a), we note that with the highest
${\cal N}=4$ proton in the unfavored $\alpha = 1/2$ [411]1/2 orbital, this
proton can be lifted to the [402]5/2 orbital as illustrated in the
figure. This will lead to two close to signature degenerate bands as
drawn in the middle panel of Fig.\ \ref{expth1314}. With this assignment,
the differences
between experiment and calculations will have almost the same spin dependence
at high spin for the three 541 bands drawn in Fig.\ \ref{expth1314} and
with a small change of the single-particle parameters placing the [402]5/2
orbital closer to the [411]1/2 orbital, the differences would be close to
overlapping.
Thus, considering energies, this appears as a convincing interpretation
but the fact that rather large configuration changes are required at
the band-crossings of the 13a,b bands makes this interpretation
somewhat questionable. However, it seems difficult to find any other
more convincing interpretation of the highest spin regions of these
541 bands (bands 13a,b).

 The decay of band 16 indicates some weak relation with the bands 13.
Furthermore, its energy curve is similar to that of these bands.
The configuration of the bands 13 at high spin is illustrated in Fig.\ \ref{sp},
i.e. with one proton in the [404]7/2 orbital. A very tentative
assignment for band 16 would then be to place this proton in the
[402]5/2 orbital instead.

\subsubsection{The transitional quadrupole moment, $Q_t$}
\label{sqt}
\begin{figure}
 \begin{centering}
\includegraphics[clip=true,width=0.48\textwidth,trim=0 0 0 0]{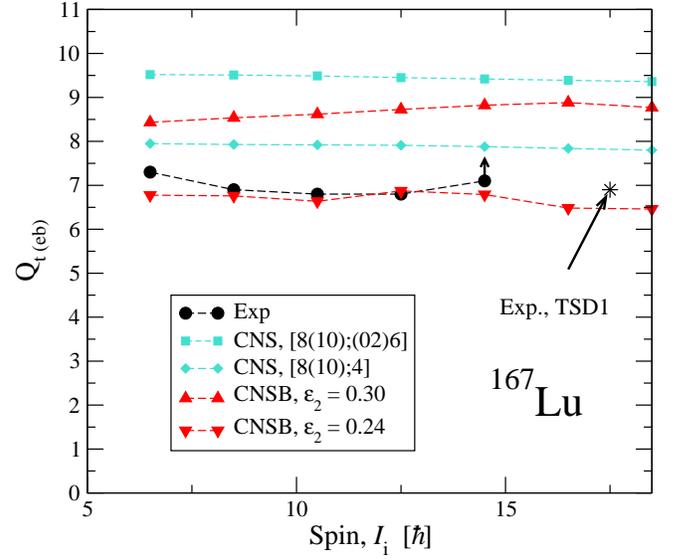} 
  \caption{\label{qt} The experimental transition quadrupole moments $Q_t$
are compared with the values calculated for fixed configurations in the CNS 
formalism; i.e. for the configuration with an [541]1/2 proton and two 
$h_{11/2}$ neutron holes, [8(10);(02)6] which is our preferred assignment,
and for the configuration with no neutron holes, [8(10);4]. Furthermore,
the values calculated in the CNSB at the two minima of the $(-,+1/2)(+,0)$
(see Fig. \ref{pes-541} are shown. Finally, the
experimental value of $Q_t$ for the TSD1 band is shown. It is obtained as
an average over the most intense transitions, see \cite{Gur05}.
}
 \end{centering}
\end{figure}

The transitional quadrupole moment of the yrast 
low-spin states up to $I=14.5$ was measured
very recently \cite{Roh19}, i.e. in the spin range where the 541 band is
yrast. The experimental values are compared with
the present calculations in Fig. \ref{qt}.
Consider first the CNS calculations, where the deformation trajectories
in Fig. 13 will give a general idea about the values of $Q_t$ 
in the different configurations. 
For our preferred configuration with two $h_{11/2}$ neutron
holes, the values
are much larger than the measured values, see Fig. \ref{qt}. Even without 
these holes, the 
calculated values are too large. 

\begin{figure}[htb!]
{\includegraphics[clip=true,width=0.25\textwidth,trim=0 28 20 0]{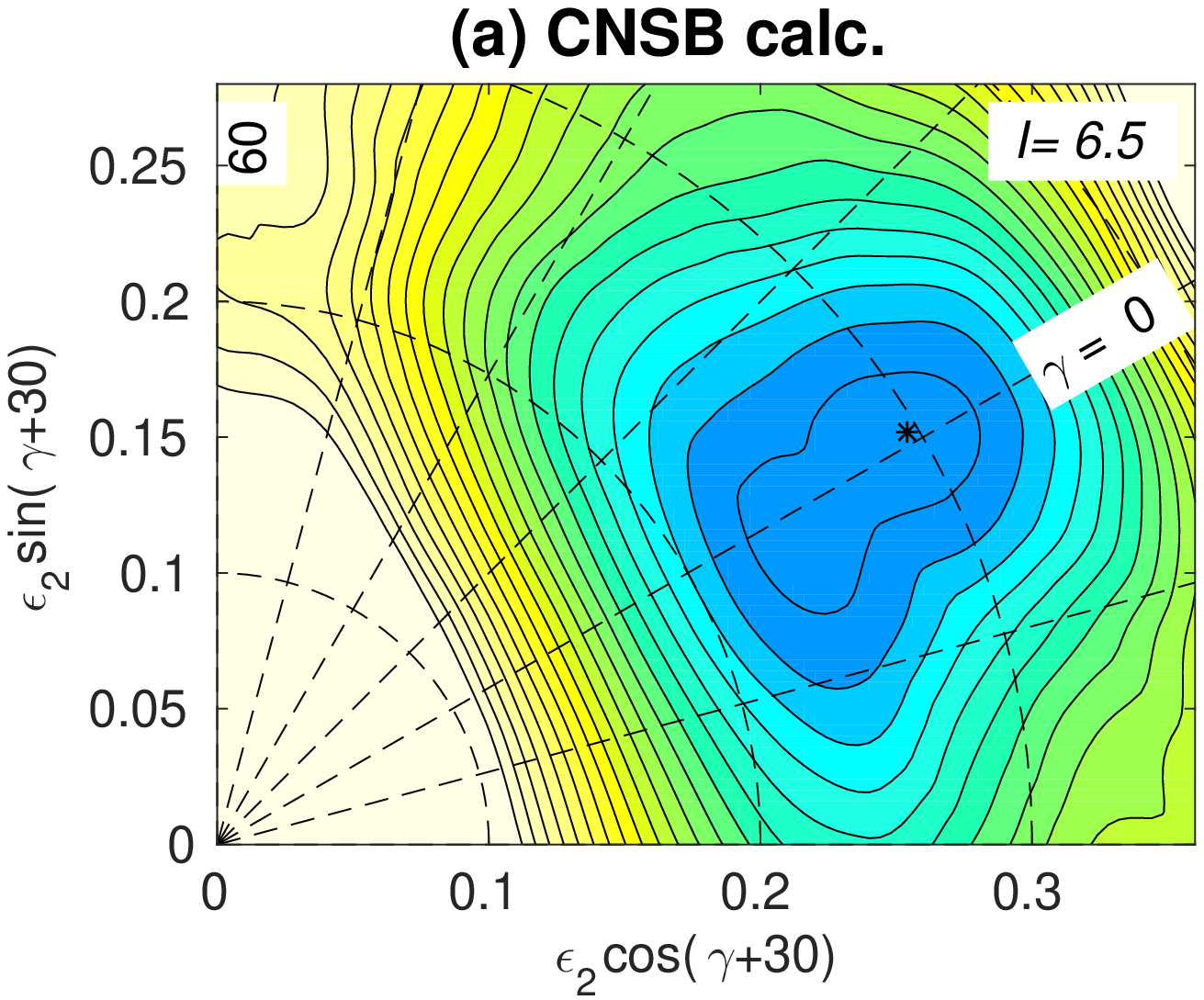}\includegraphics[clip=true,width=0.23\textwidth,trim=30 28 20 0]{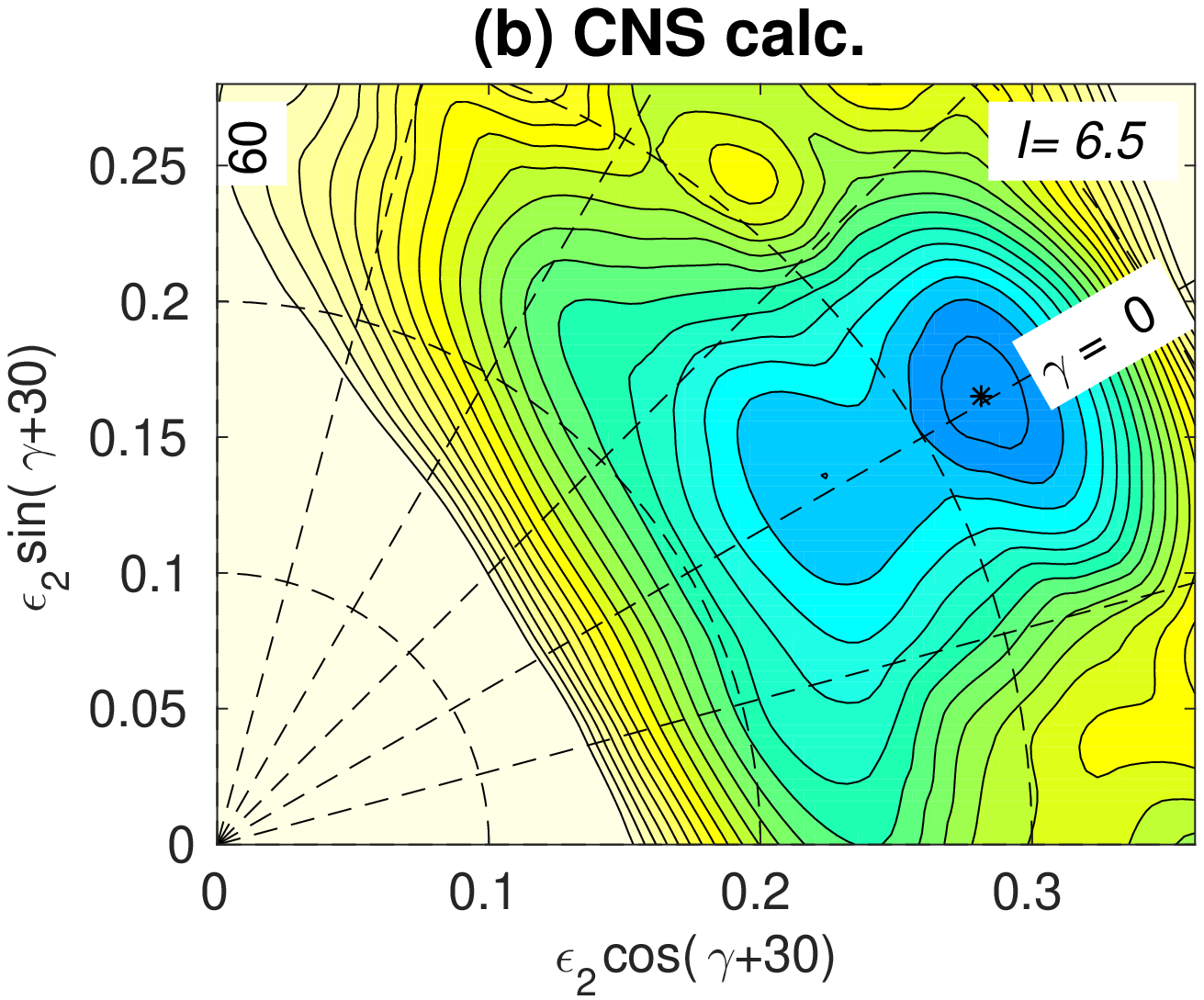}}
{\includegraphics[clip=true,width=0.25\textwidth,trim=0 28 20 0]{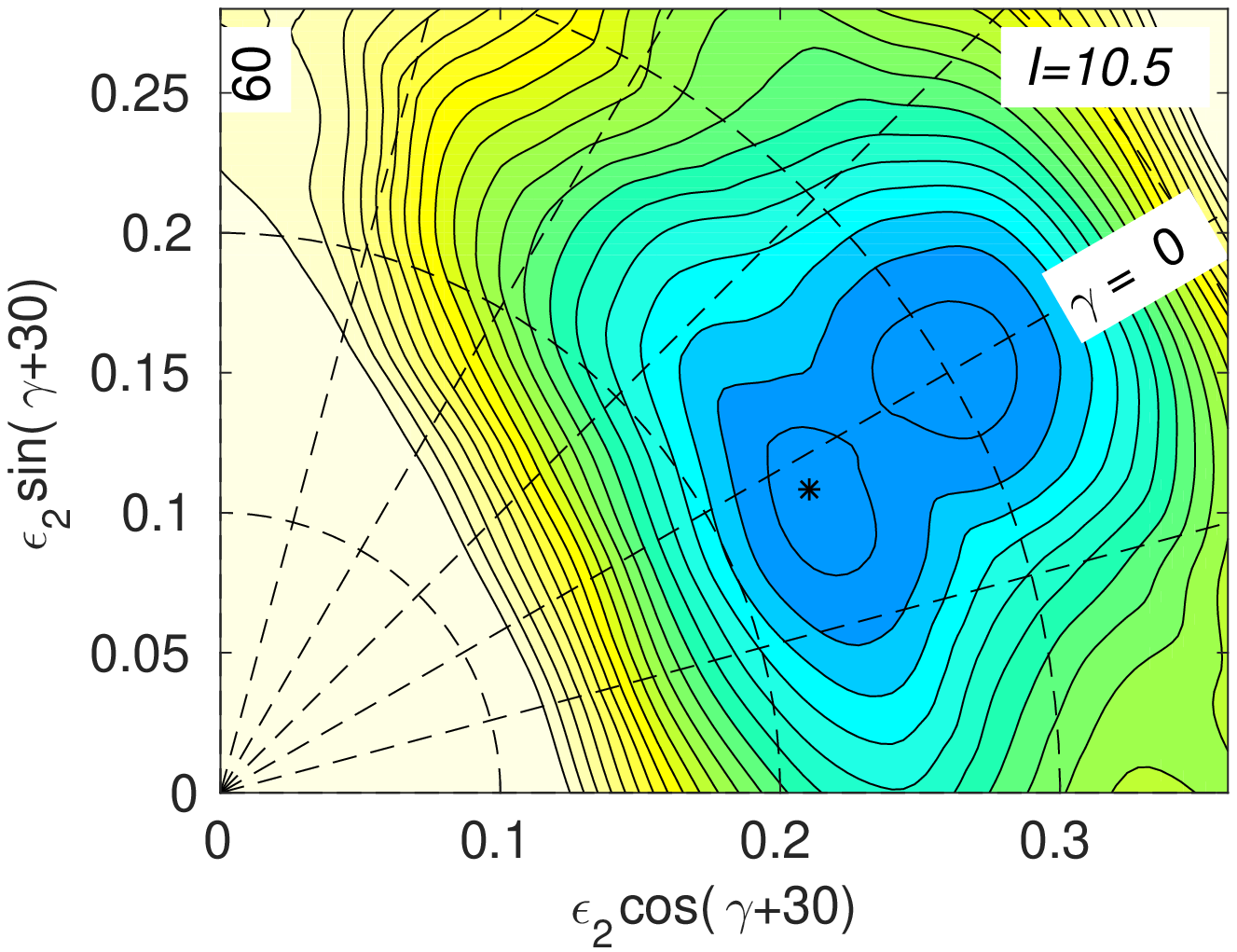}\includegraphics[clip=true,width=0.23\textwidth,trim=30 28 20 0]{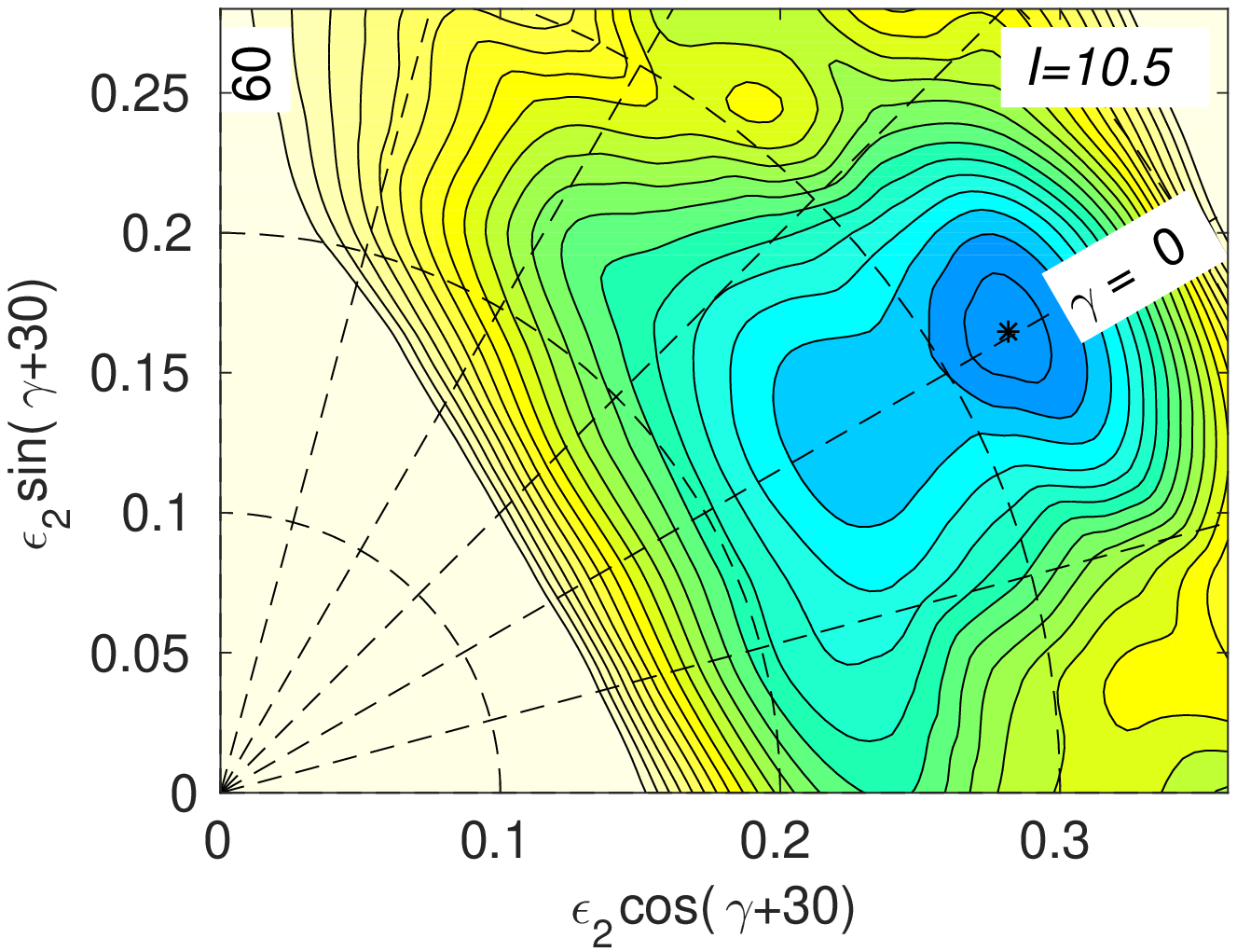}}
{\includegraphics[clip=true,width=0.25\textwidth,trim=0 0 20 0]{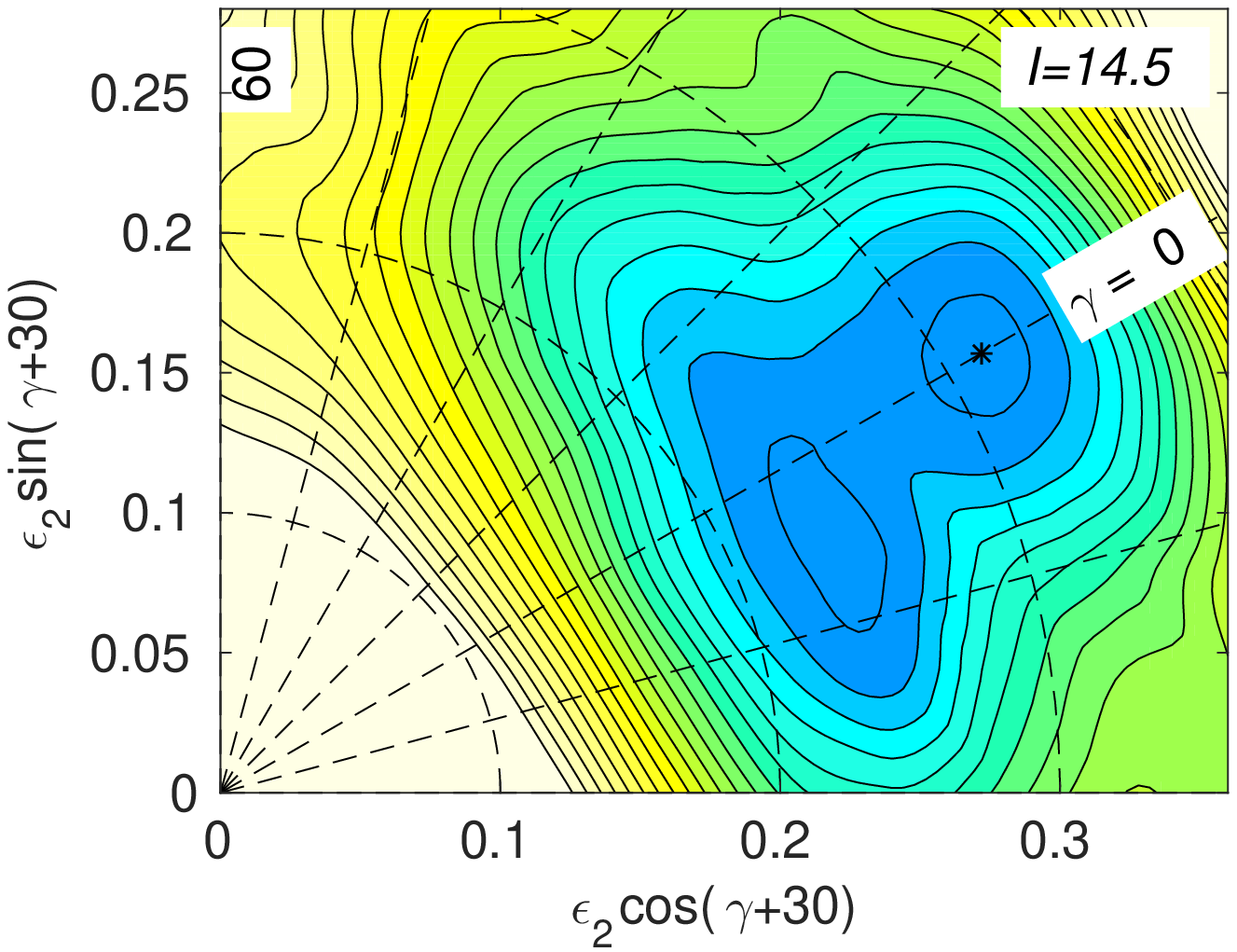}\includegraphics[clip=true,width=0.23\textwidth,trim=30 0 20 0]{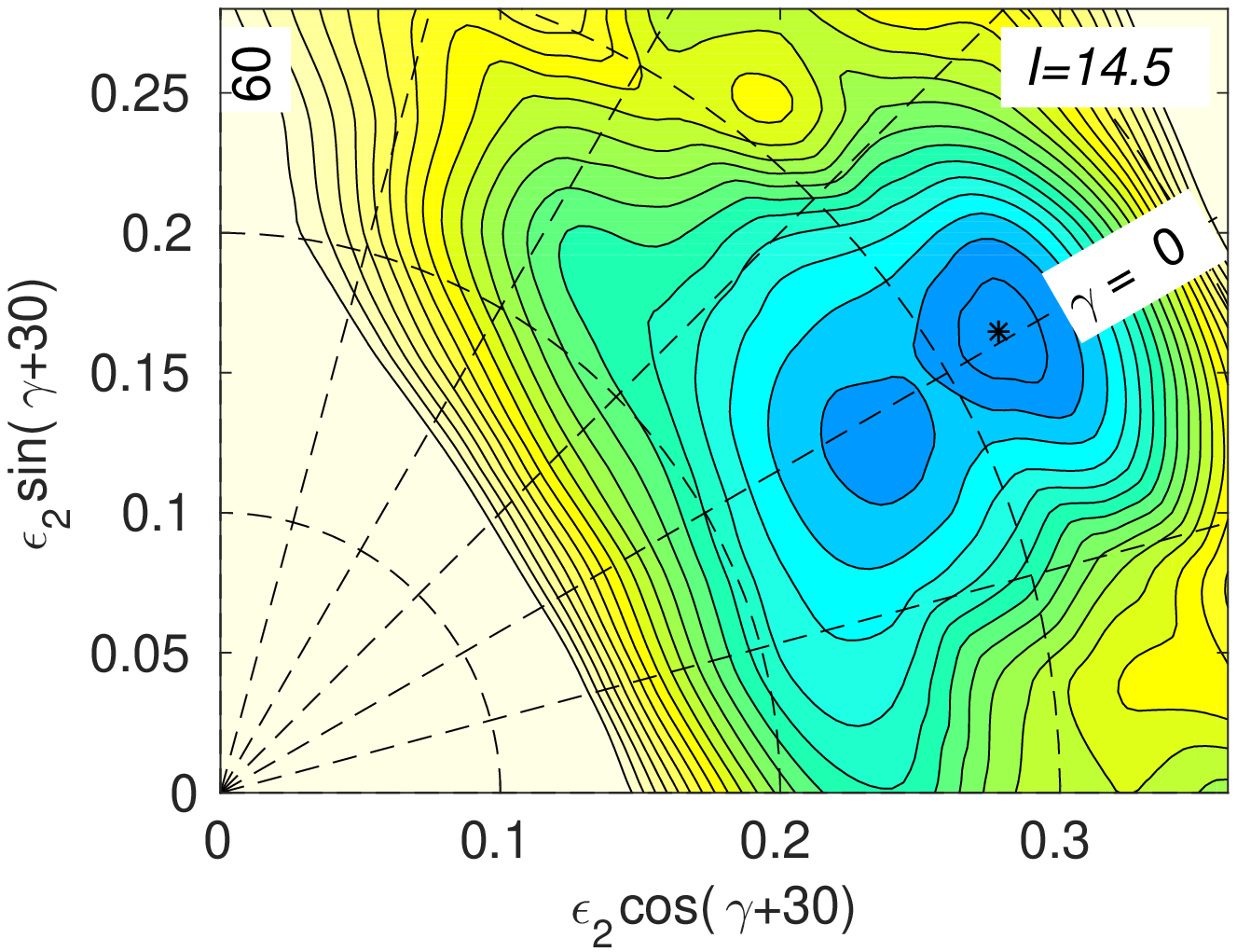}}
\caption{Calculated total energy surfaces in the 
($\varepsilon_2, \gamma$)-plane (a) with pairing included (CNSB) and (b) with no pairing 
(CNS) for the
$(-,+1/2)(+,0)$ configuration at spin values $I=6.5, 10.5$ and 14.5. In each
mesh point, the energy has been minimized with respect to $\varepsilon_4$
deformations. The contour line separation is 0.2 MeV.}
\label{pes-541}
\end{figure}

In order to get a better understanding of
the different configurations, the corresponding energy surfaces are drawn
in the CNS as well as in the CNSB formalism for spin values $I = 6.5, 10.5$ and
14.5 in Fig. \ref{pes-541}. 
There are two minima in these surfaces, one at 
$\varepsilon_2 \approx 0.30$ and the other at  $\varepsilon_2 \approx 0.24$
where the former corresponds to our preferred configuration for band 13
at low spin, i.e. with a [541]1/2 proton and two 
$h_{11/2}$ neutron holes, while the latter corresponds to the configuration
for the $h_{11/2}$ band, i.e. with no neutron holes and an odd $h_{11/2}$
proton. In the CNS calculations, the former configuration is clearly lowest
in energy while the two configurations come at a similar energy in CNSB.
This shows that the pairing energy is unusually strong in 
configurations with the odd proton in an $h_{11/2}$ orbital as noticed
previously in Secs.\ \ref{comp} and \ref{cnsb}. 

The comparison between CNS and CNSB
calculations for the $\varepsilon_2 \approx 0.30$ minimum shows 
that the deformation
of this configuration is clearly smaller with pairing included, illustrating
the general effect of pairing that it tends to reduce the shell effects,
see e.g. Fig. 14.5 of Ref.\ \cite{Nil95}. Coming back to Fig.\
\ref{qt}, the calculated $Q_t$ value for the large deformation minimum 
is thus 
considerably smaller with pairing included but still much larger than 
experiment. On the other hand, the calculated value of $Q_t$ for the 
small deformation minimum comes close to experiment. However, this
minimum corresponds to the $h_{11/2}$ configuration and thus not to 
the [541]1/2 configuration which is assigned to negative parity yrast
band at low spin. Coming back to this [541]1/2 configuration with
no holes, its $Q_t$ value calculated in CNS is larger than experiment
but still reasonably close, see Fig. \ref{qt}. As it does not show 
up as a local minimum in the
energy surfaces of Fig.\ \ref{pes-541}, we cannot 
easily calculate its value with
pairing included. However if the value is somewhat reduced by pairing in
a similar way as for the $\pi([541]1/2)\nu(h_{11/2})^{-2}$ configuration, 
it will
come close to experiment. On the other hand, the way that bands 13a
and band 14 cross at $I \approx 18$ strongly suggests that they have
different configurations and thus that yrast band where $Q_t$ has
been measured, band 13a, has two $i_{13/2}$ holes for spin values
below $I \approx 22$.
The present calculations can be compared with the TRS calculations
presented in Ref. \cite{Roh19}. They were done without keeping
track of configurations. The fact that 
the calculated values came pretty close to the experiment 
indicates that the TRS yrast configuration has no $h_{11/2}$ holes.

The measured
value for the band labeled as TSD1 is also shown in Fig. \ref{qt}. The figure
suggests that the values of $Q_t$ for the TSD1 band and the 541 band are very
similar. This appears strange as has been noted before, see \cite{Rag17}. One
should note that the experimental values have an uncertainty of approximately
15 \% because of systematic errors. Therefore, it would be important to measure
$Q_t$ of the different normal-deformed and TSD bands in the same experiment in 
which case it should be possible to determine the relative values with a
much better precision.


\subsection{Three-quasiparticle (3qp) bands}
\label{sect-3qp}
\subsubsection{The observed BC crossing in the 3qp configurations}
\begin{figure}
 \begin{centering}
\includegraphics[clip=true,width=0.45\textwidth,trim=0 0 0 0]{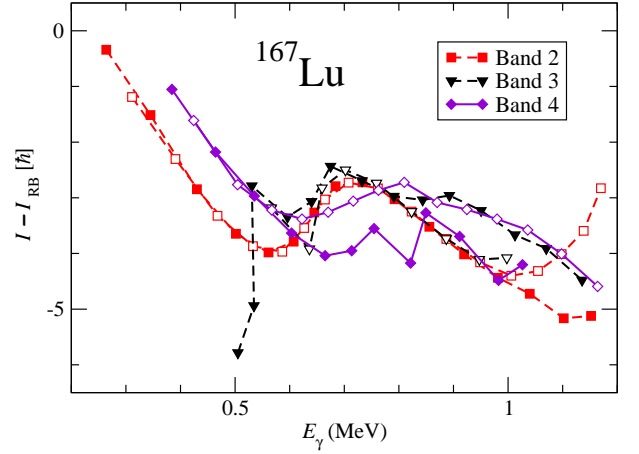} 
  \caption{\label{ivseg-3qp} The spin value with a rigid-body reference subtracted
for the three-quasiparticle bands is shown versus the transition energy, 
$E_{\gamma}$.}
 \end{centering}
\end{figure}

The experimental bands 2, 3 and 4 are assigned as built on two
quasi-neutrons and one quasi-proton in Ref.\ \cite{rou15}. A smooth 
crossing is
observed in band 2 at the frequency $\hbar \omega = 0.32$ MeV with
an alignment of $2-3 \hbar$. This is discussed
in the appendix where also plots of $I$ vs. $E_{\gamma}$ with a
rigid body reference subtracted were introduced. The bands
2, 3 and 4 are shown in such a diagram in Fig.\ \ref{ivseg-3qp}.
It is evident that all these bands go through a similar crossing,
where the alignment is rather somewhat smaller for bands 3 and 4
and the crossing frequency (or $E_{\gamma}$) is somewhat larger in
band 4 than in bands 2 and 3.  
%
The presence of this crossing, the  $i_{13/2}$ BC crossing, and the absence
of the AB crossing at a lower frequency shows that
these bands have an odd number of  $i_{13/2}$ neutrons,
i.e. they can be assigned as three-quasiparticle configurations
in agreement with Ref.\ \cite{rou15}.
We note that while the alignment in the AB crossing is quite large of the
order of $8 \hbar$, it is much smaller in the BC crossing, namely of the
order $2 \hbar$. This is understood from the fact that one $i_{13/2}$ neutron
is already aligned (and `blocked') in configurations which go through the
BC crossing. The next crossing would then be the CD crossing in configurations
with two aligned  $i_{13/2}$ neutrons. Thus, this crossing is expected to be
much less distinct that the BC crossing and it seems natural that it is not
seen at all, in agreement with our conclusions in Sec.\ \ref{n4}.  


\subsubsection{Configurations of the 3qp bands which are calculated low in energy}
\begin{figure}
 \begin{center}
\includegraphics[clip=true,width=0.48\textwidth,trim=0 0 0 0]{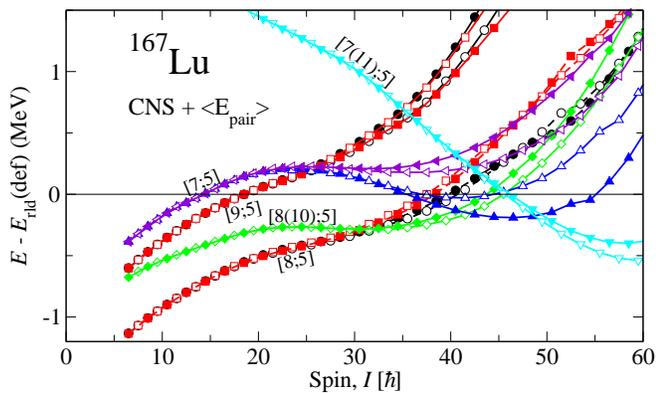}
  \caption{\label{3qp} Calculated `CNS + average pairing'
  energies for low-lying bands in
    $^{167}$Lu with 9 $hf$ and 5 $i_{13/2}$ neutrons.}
 \end{center}
\end{figure} 
The configurations of the favored three-quasiparticle bands can be
read out from Fig.\ \ref{sp}. Thus the neutron $(i_{13/2})^5$
configuration is combined with proton configurations with the
odd particle in ${\cal N}=4$ orbitals, in $h_{11/2}$, or in the $hf$ orbital
[541]1/2. The lowest calculated configurations 
of this type are drawn relative
the rotating liquid drop reference in Fig.\ \ref{3qp}. They are
calculated in the CNS approach with an average pairing added. 
For spin values below $I \approx 30$, the  
lowest configuration has 8 $h_{11/2}$ protons and thus the
odd proton in the [404]7/2 orbital. The orbitals of both the odd
proton and the odd ${\cal N}=5$ neutron are signature degenerate up to high
frequencies so four bands, which are essentially overlapping in Fig.\ \ref{3qp} up to $I \approx 30$,
are formed. 
The calculated yrast states for $I \approx 30-40$ have the odd proton
in the [541]1/2 orbital, where for the two bands drawn in Fig.\ \ref{3qp},
the favored $\alpha = 1/2$ signature of the [541]1/2 proton
is combined 
with both signatures for the $hf$ neutron. 
Bands with an
odd number of $h_{11/2}$ protons are also calculated low in energy,
where those with the odd
proton in [523]7/2 (configuration [7;5])  and  in [514]9/2 (configuration [9;5]) 
come at a similar energy at low spin as would be expected from
Fig.\ \ref{sp}, while the configuration with fewer $h_{11/2}$ protons
is clearly favored at high spin. Indeed, it is this [7;5]
configuration which is assigned to the observed 
positive parity band labeled
404h, which is yrast above $I \approx 30$. Finally,
in Fig.\ \ref{3qp} we have also added the lowest configuration
with one proton in the $i_{13/2}$ orbital, [7(11);5].
Note that this configuration should be classified as having at
least five quasiparticles because it has an odd proton in the
$h_{11/2}$, $hf$ and $i_{13/2}$ orbitals in addition to the two odd
neutrons.
Configurations of this type are calculated to become yrast 
around $I=50$ but because the 
position of the $i_{13/2}$ shell is not well established,
this number is very
uncertain. Several of the observed bands show 
discontinuities at the highest spin values. One possibility
is that these  discontinuities are caused by 
crossings with bands with one $i_{13/2}$ proton.

\subsubsection{Comparison between observed and calculated 3qp configurations}
\begin{figure}
 \begin{center}
\includegraphics[clip=true,width=0.4\textwidth,trim=0 0 0 0]{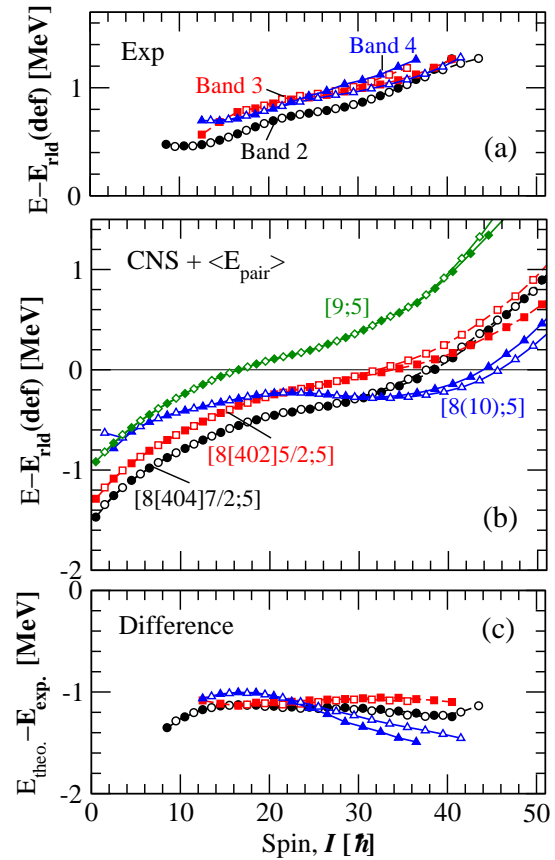}
  \caption{\label{expth23} Experimental energies (a) and theoretical 
`CNS + average pairing' energies (b) relative to a rotating liquid drop and their differences (c) as a function of spin
 for bands 2, 3 and 4 in $^{167}$Lu.}
 \end{center}
\end{figure}

The observed and calculated three-quasiparticle configurations
are compared in Fig.\ \ref{expth23}. The bands 2 and 3 could
be assigned to the four [8;5] configurations drawn in Fig.\ \ref{3qp}.
However, there is also the possibility to form excited CNS bands,
i.e.\ the second or third lowest band within some CNS configurations.
Such configurations were previously considered for the ${\cal N}=4$
positive parity bands. In an analogous way, in the [8;5] bands
drawn in Fig.\ \ref{3qp}, the odd proton is placed in the [404]7/2
orbital but it can be lifted to the [402]5/2 orbital at a low
energy cost. Thus, the two lowest bands with the odd proton in
[404]7/2 and [402]5/2, respectively, are drawn in Fig.\ \ref{expth23}
and compared with the observed bands 2 and 3. The difference
between calculations comes out as rather constant around $-1$ MeV
which clearly supports this interpretation. The fact that the 
observed bands are split by approximately 200 keV at low spin
makes this interpretation more plausible than the assumption that
bands 2 and 3 are built on the four [8;5] bands shown in 
Fig.\ \ref{3qp}. On the other hand, one should also note that
there will be some residual interaction between the bands,
i.e.\ the observed bands should rather be assigned to some
mixture of the pure CNS bands. Such mixing, however, will
only have some minor influence on the energies and will
not really be noticed on the energy scales used in Fig.\ \ref{expth23}.  

The interpretation of bands 2 and 3 as [8;5] is further supported by
the comparison with the corresponding paired configuration,
i.e.\ the lowest $(+,\pm1/2)(-,1)$ in Fig.\ \ref{expth2}. Note that the 
difference curve for band 2 comes close to that for the
$h_{11/2}$ band and that the spin dependence for the two difference
curves in the lower panel of Fig.\ \ref{expth2} is very similar.


As seen in Fig.\ \ref{3qp}, among three-quasiparticle
configurations with positive parity, [8(10);5] is the lowest
one for spins $I<40$. This suggests that band 4 should be assigned 
to this configuration. One might also argue that
 the crossing for signature
$\alpha = -1/2$ between band 4 and bands 2 and 3 at spin $I \approx 30$ is
reproduced for the configurations [8(10);5] and [8;5]. 
However, considering the overall features, it is evident that
the observed bands 2, 3 and 4 have a rather similar spin dependence
while for the calculated bands, the [8(10);5] configuration comes
down a few hundred keV relative to [8;5] in the spin range
$I \approx 15-40$.
In Fig.\ \ref{3qp}, it appears that the energy curve for the [9;5]
configuration is more parallel to that of the [8;5] configuration
in general agreement with the behavior of bands 2, 3 and band 4. 
This would suggest that band 4 should rather be assigned to the [9;5] 
configuration as done on Ref.\ \cite{rou15}. 
However, considering the differences in detail, it turns out that the
calculated energy curve of the configuration [9;5] comes up too
steep at high spin leading to a difference between calculations
and experiment (not shown in Fig.\ \ref{expth23}) which increases
in an unrealistic way. Thus, the assignment of band 4 to the
[8(10);5] configuration is well motivated.

A final observation for the three-quasiparticle bands is that one would 
expect to observe the [7;5] configuration because in Fig.\ \ref{3qp}, it is 
clearly calculated lowest in energy for $I=40-50$. It has no 
correspondence among the observed low-lying bands labeled as 3qp,
but instead, in section \ref{n4}, it was assigned to the positive parity 
band which is yrast for $I \approx 30-50$, the 404h band. The fact that
no other observed band is naturally assigned to the [7;5] configuration
gives a strong support to our assignment for the 404h band.

\section{Summary and conclusions}
\label{summ}
We have employed the unpaired CNS and paired CNSB formalisms to
determine the structure of the observed rotational bands in
$^{167}$Lu. It turns out that
the configurations 
calculated with
pairing included are well described in the unpaired formalism if
an average pairing energy is added. Thus, the more detailed 
configuration assignment in the unpaired formalism in terms 
of number of particles in
different $j$-shells or groups of $j$-shells can be used to
classify the observed bands.
%
This is possible even though some features show up only
in the paired formalism, especially the low energy in the 
$(\pi,\alpha) = (+,0)$ neutron configurations
below the first $(i_{13/2})^2$ backbend at  $I=12-14$, the AB crossing. 
Also in configurations
with an odd number of  $i_{13/2}$ neutrons, the paired  $(i_{13/2})^2$
 crossing, the BC crossing, is seen for spin values $I=15-20$, but the alignment is so small that its
contribution to the total energy is almost negligible. 
Furthermore, the
full pairing calculation is important in an extended spin range
for one-quasiparticle configurations 
with the odd proton in an 
orbital far away from the Fermi surface.


Some positive parity bands
which interact are redefined into structures which
evolve smoothly with spin, where approximate interaction strengths
between these structures are extracted.  
%
Some previous configuration assignments are
confirmed, while the interpretation of some experimental bands and
also the origin of the crossings are revised. For $^{167}$Lu with an
odd number of protons, the low spin one-quasiparticle configurations
are followed at somewhat higher spin values by three-quasiparticle
configurations with an odd number of $hf$ and $i_{13/2}$ neutrons.
Note however that at high spin, say $I>30$, the division into configurations 
with a fixed number of quasiparticles is not meaningful because, 
with weaker pairing correlations, the different
spin vectors will align gradually, i.e.\ the spin vectors of all
valence particles will be partially aligned.

In our interpretation, those configurations which would be expected to come
low in energy according to a single-routhian diagram, see Fig.\ \ref{sp},
have all been localized in the observed level scheme. Similarly,
reasonable interpretations have been found to all low-lying
bands which are observed in an extended spin range.

Considering our preferred assignments,
the extension of the [404]7/2 band after the band-crossing at
$I \approx 30$ is assigned
as a three-quasiparticle configuration with negative parity for
both protons and neutrons, while the smooth continuation of the 
[404]7/2 band is observed as non-yrast up to $I \approx 40$. 
In the observed $h_{11/2}$ band,
the band-crossing at $I=40$ is caused by the occupation of the
lowest $h_{9/2}$ orbital, [541]1/2. This $h_{9/2}$ orbital is 
assigned to the band-head of a one-quasiparticle configuration
which comes low in energy because of an increased deformation caused
by two $h_{11/2}$ neutron holes. In this band, there is a band-crossing at 
$I \approx 20$ where the configuration with no $h_{11/2}$ holes
comes lower in energy. The positive parity three-quasiparticle
band is assigned as built on a configuration with one $h_{9/2}$
proton. In general, the relative energy of bands assigned as having one 
$h_{9/2}$ proton and no $h_{9/2}$ proton, respectively, agrees
with experiment. This fact indicates
that the proton $h_{9/2}$ shell situated above the $Z=82$ gap is placed 
at a proper energy with present single-particle parameters.  

%
%

We have not tried to interpret the band which is assigned as TSD,
i.e.\ `triaxial strongly deformed'. 
However, we have noticed the strange feature
that this TSD band interacts strongly with one or several normal-deformed
bands at $I \approx 30$ and tried to extract the interaction 
strength. A problem is however that bands which appear to interact
with the TSD band are observed only for spin values below the
interaction region. It would thus be important to follow these
bands in experiment through the interaction region. 

Measurements of the life-times in the TSD band 
and the yrast ND band have been presented \cite{Gur05,Roh19}. 
However, these life-times are associated with large uncertainties
and appear somewhat confusing when compared with
the present calculations. Thus, it would be important to carry
out life-time experiment for several bands including the TSD band
in the same experiment, which should make it possible to determine
at least the relative life-times with a much better accuracy.
 
Our calculations are limited to one nucleus but should be valid 
in general not only for
the rotational bands in the deformed rare-earth region but also
for high-spin states in other mass regions. This concerns for example
the observation that `paired band crossings' are mainly seen at
low spin and absent in the very high spin region, 
and the conclusion that after the first or possibly second
band crossing, high-spin states should not be 
classified by the number of quasi-particles but rather by the 
distribution of particles over the different $j$-shells or groups of
$j$-shells. The present study is the first one where the CNS and CNSB
methods have been combined to describe a large number of rotational
bands for a nucleus in the deformed rare-earth region, but previous
studies of the transitional nucleus $^{161}$Lu \cite{ma14} 
or on the yrast bands in $^{168-175}$Hf \cite{tah18} support 
that our conclusions can be considered to be a more general nature.

\appendix
\section{Band crossings and aligned spin}
\begin{figure}
 \begin{centering}
\includegraphics[clip=true,width=0.49\textwidth,trim=0 0 0 0]{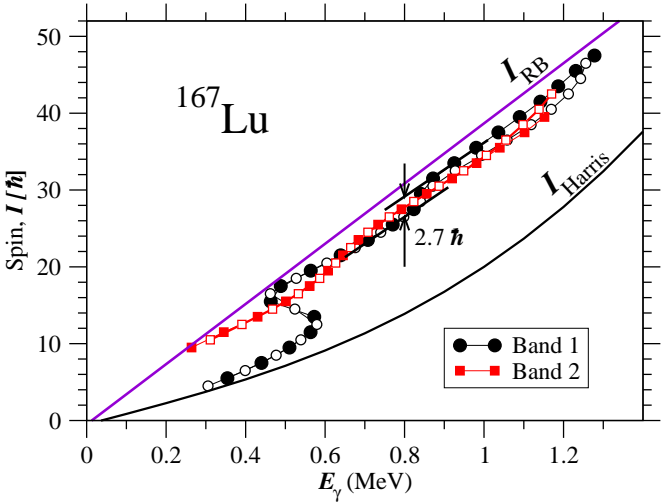}
  \caption{\label{align-b12} Spin $I$ versus transition energy 
$E_{\gamma}$ for the bands labeled
1 and 2 in Ref.\ \cite{rou15}, see Fig.\ 17 in that reference. 
Furthermore, the rigid body
spin, $I_{RB}$, and the spin according to Harris' formula \cite{Har65},
$I_{Harris} = \omega {\cal J}_0 + \omega^3{\cal J}_1$
(${\cal J}_0 = 27 \hbar^2 {\rm MeV}^{-1}$, ${\cal J}_1 = 56 \hbar^4 {\rm MeV}^{-3}$)
are drawn.} 
\end{centering}
\end{figure}

Much of the analysis of high-spin bands has been based on band-crossings,
where these crossings are often assumed to be caused by pairing. 
For such crossings, the alignment is an important
quantity when comparing experiment and calculations. 
However, different definitions of the alignment have been used in the
litterature.
Especially, it
appears that the definition used in Ref. \cite{rou15} is different from
the one we use.
In view of this, we will give some comments on how band crossings can
be analyzed and consider in some detail the crossings seen in the
bands labeled as 1 and 2 in Ref.\ \cite{rou15}. Thus, in
Fig.\ \ref{align-b12} the spin $I$ of these bands is plotted versus
the transition energy, $E_{\gamma}$. Note that we define the
rotational frequency as $\hbar \omega = E_{\gamma}/2$, i.e.\ 
$\hbar \omega$ is half the transition energy, so it is equivalent to plot $I$
vs. $\omega$. Furthermore, because we are mainly interested in high spin
states, we identify the total spin with its projection on the rotation
axis, $I \equiv I_x$. 

Let us first concentrate on the second crossing in the 
$\alpha=+1/2$ branch of band 1, i.e.\ the crossing which was analyzed in
connection with Fig.\ \ref{exp1}(b). In order to find out how much
spin is gained in this crossing, we draw two parallel lines following
the transitions below and above the crossing and can conclude that the
lines are displaced by approximately $2.7 \hbar$. The conclusion is
thus that in addition to the smooth spin increase from the core, $2.7
\hbar$ is gained from the change of the wave-function at the
crossing. This change could either be caused by the alignment of a pair of
high-$j$ particles or a change of the orbital occupation.

\begin{figure}
 \begin{centering}
\includegraphics[clip=true,width=0.45\textwidth,trim=0 0 0 0]{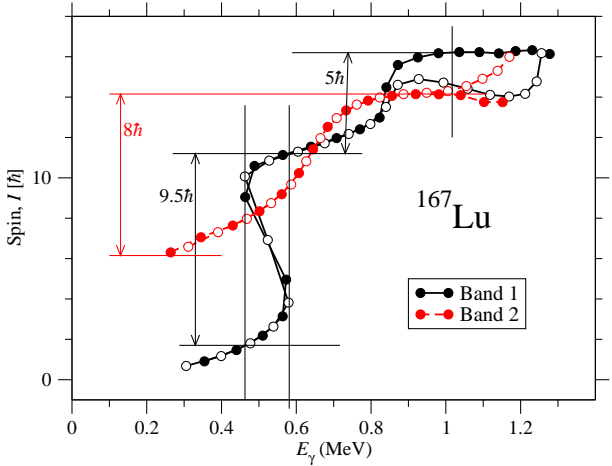}
  \caption{\label{harris} 
The spin $I$ relative to $I_{Harris}$ with ${\cal J}_0 = 27 \hbar^2 {\rm MeV}^{-1}$ 
and ${\cal J}_1 = 56 \hbar^4 {\rm MeV}^{-3}$ for the same bands as in Fig.\ \ref{align-b12}. 
Lines are drawn in to show how the alignments listed in Table III of
Ref.\ \cite{rou15} can be obtained.
}
 \end{centering}
\end{figure}
In the analysis of band crossings,
a reference spin 
$I_{ref}$ is often
subtracted from the total spin where this reference is generally chosen from
the parameterization of Harris \cite{Har65}, 
$I_{ref} = \omega {\cal J}_0 + \omega^3{\cal J}_1$. 
The idea is then to choose the constants ${\cal J}_0$
and ${\cal J}_1$ such that the reference spin follows the smooth raise
of the $I$ vs. $\omega$ (or $I$ vs. $E_{\gamma}$) curve before and
after the crossing, i.e. $I - I_{ref}$ will be approximately constant
outside the band crossing region \cite{Ben79}. 
However, it is in general difficult
to find constants ${\cal J}_0$ and ${\cal J}_1$ so that this constancy
is fulfilled in the entire spin range.
Consider e.g. the constants
chosen in Ref. \cite{rou15} where the corresponding reference spin,
$I_{Harris}$, is shown in Fig.\ \ref{align-b12} while the
$I-I_{Harris}$ curves for bands 1 and 2 are shown in Fig.\ \ref{harris}.  
It appears that in this case, the
constants have been chosen such that the  $I - I_{ref}$ curves become 
constant at high transition energies, $E_{\gamma} \approx 0.8-1.2$ MeV. 
Then,  even though
the curves are not constant outside band crossing regions at smaller 
values of $E_{\gamma}$, it is assumed that $I_{ref}$   
represents the smooth
average increase of the spin, i.e. that the full spin of the  
$I - I_{ref}$ curve
is built from alignments at band-crossings. 
There is still some arbitrariness about over which $E_{\gamma}$ range the
different alignments should be counted. However, if it is assumed that 
for a full backbend, only
the frequency region (or $E_{\gamma}$ region)  covering the backbend should
be  counted, the alignment at the different crossings of the 
$\alpha = 1/2$ bands can be extracted using the lines drawn in
Fig.\ \ref{harris}. Thus, 
alignments of $9.5 \hbar$ and $5 \hbar$ can be read out
from the first and second band crossings in band 1 and $8 \hbar$ for
the crossing at $E_{\gamma} \approx 0.64$ MeV in band 2. 
 
A problem with the definition of the  spin alignment used in Ref. 
\cite{rou15} is that it will depend on the $I_{ref}$ curve which is
subtracted. For example, in the previous study of high spin states in 
$^{167}$Lu \cite{Yu90}, the constants  ${\cal J}_0$ 
and ${\cal J}_1$ were chosen
such that the $I - I_{ref}$ curve became essentially constant before
and after the first band-crossing of band 1. The alignment 
at  this crossing was then extracted as $8.5 \hbar$. This
value is consistent with the value we get if 
parallel lines are drawn in Fig.\ \ref{align-b12} in a similar
way as for the second crossing.
The conclusion is that only if an  $I_{ref}$ curve can be defined
which makes $I - I_{ref}$ essentially constant in some $E_{\gamma}$ (or $\omega$)
region both before and after the crossing, this is consistent way
to define the aligned spin.  

\begin{figure}
 \begin{centering}
\includegraphics[clip=true,width=0.4\textwidth,trim=0 0 0 0]{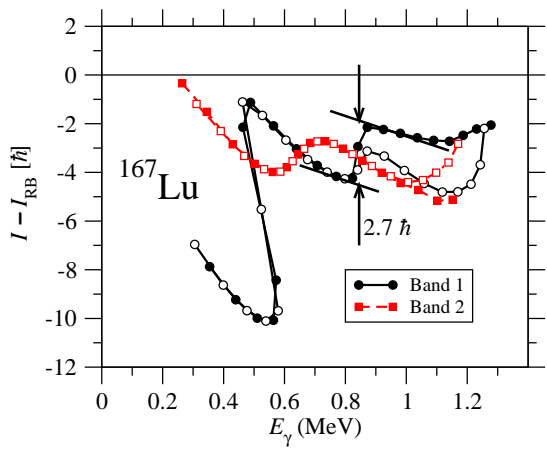} 
  \caption{\label{rb} The spin relative to $I_{RB}$ is drawn vs. $E_{\gamma}$ for 
the same bands as in Figs.\ \ref{align-b12} and \ref{harris}. In an analogous way
as in Fig.\ \ref{align-b12}, lines are drawn to indicate how an alignment of $2.7 \hbar$
can be obtained for the second crossing in band 1.}
 \end{centering}
\end{figure}

It can be helpful to subtract some reference from the $I$ vs. 
$E_{\gamma}$ curve to see the different discontinuities more clearly.
When considering the full spin range up to $I \approx 50$, 
it is evident that the spin $I$ is roughly 
proportional to $E_{\gamma}$, see Fig.\ \ref{align-b12}. 
Then, the strategy is to choose the reference 
such that the differences from this straight line behavior
are highlighted. 
Different possibilities which in practice are very similar are
to choose ${\cal J}_1 = 0$ in the Harris formula
or to use some reference of rigid
body type, where the latter option appears more natural in 
connection with the present CNS and CNSB calculations. 
The behavior will be similar if the 
rigid body moment of inertia is calculated
at a fixed deformation 
or alternatively
calculated at the minimum energy of the rotating liquid drop energy
as in our standard reference energy \cite{carl06}, 
see e.g.\ Figs.\ \ref{exp} and \ref{exp1}.
With the rigid body constants chosen according to Eq. (70) in Ref.\ \cite{Afa99},
$I_{RB}$ comes out as in Fig.\ \ref{align-b12}, while
 $I - I_{RB}$ is drawn versus 
$E_{\gamma}$ for bands 1 and 2 in Fig.\ \ref{rb}. 
It is now easy to read out that the alignment at
the second crossing of Band 1 is approximately $2.5 \hbar$. If the parallel lines are
drawn as in Fig.\ \ref{align-b12}, i.e.\ following the trend of a few transitions before and
after the crossing, we will of course get the same answer, $2.7 \hbar$,
according to the lines drawn.
If the points closest to the crossing is given a higher weight, i.e.\ lines which come
closer to a constant value of $I - I_{RB}$, we will get somewhat smaller values of the
alignment, down to $\approx 2.2 - 2.3 \hbar$. Thus. it is difficult to define a very precise
value of the alignment. On the other hand, a value which is precise enough for the
theoretical analysis can easily be deduced either from the $I$ vs. $E_{\gamma}$ plot or 
from the plot with $I_{RB}$ subtracted from the spin $I$. 

Let us come back to the small alignment of
band 2 at $E_{\gamma} \approx 0.65$ MeV. This alignment is almost
invisible in the $I$ vs. $E_{\gamma}$ plot
in Fig.\ \ref{align-b12}.
However, it shows up clearly in  Fig.\ \ref{rb} 
when $I_{RB}$ is subtracted, where one can easily get a rough estimate of 
approximately $2 \hbar$ for the
alignment. Note that in this case, an alignment of $8 \hbar$ was deduced in Ref. \cite{rou15},
see Fig.\ \ref{harris}. Note also that it is not easy to extract the
alignment at this crossing according to our definitions from 
Fig.\ \ref{harris}.
It appears that this is because of the $\omega^3$ term which results in a curvature  
in the $I_{Harris}$ function. Especially, going to very high spins, the $\omega^3$ term will
dominate making this reference unrealistic. 

Summarizing, it is evident that depending on definitions, 
very different values can be deduced for the alignment at a band-crossing. However,
if we accept that
the alignment should measure the additional spin which is gained at a band-crossing,
the uncertainties become much smaller and 
the value can easily be determined with an acceptable accuracy. For example, in order 
to extract the alignment for the first back-bending, the method to subtract a reference,
$I_{ref}$ based on Harris formula will give well-defined values as long as the parameters
of the reference are fitted for this specific spin range. However, in order to extract
alignments at different spin values up to high spins, a reference of rigid body type
appears more suitable. 


\begin{thebibliography}{15}
\bibitem{Joh72} A. Johnson, H. Ryde, and S. Hjorth, Nucl. Phys. A {\bf 179}, 753 (1972).
\bibitem{Ste75} F. S. Stephens, Rev. Mod. Phys. {\bf 47}, 43 (1975).
\bibitem{Boh78} A. Bohr and B. Mottelson, Proc. Int. Conf. nuclear structure, Tokyo, Sept. 5-10, 1977, J. Phys . Sac.
Japan 44 (1978), suppl. p. 157.
\bibitem{Ben79} R. Bengtsson and S. Frauendorf, Nucl. Phys. A {\bf 314}, 27 (1979); Nucl. Phys. A {\bf 327}, 139 (1979).
\bibitem{Rie81} L.L. Riedinger, Phys. Scr. {\bf 24}, 312 (1981).
\bibitem{Voi83} M. J. A. de Voigt, J. Dudek, and Z. Szymański Rev. Mod. Phys. {\bf 55}, 949 (1983).
\bibitem{beng85} T. Bengtsson and I. Ragnarsson, Nucl. Phys. A {\bf 436}, 14 (1985).
\bibitem{Rag95} I. Ragnarsson, V.P. Janzen, D.B. Fossan, N.C. Schmeing, and R. Wadsworth, Phys. Rev. Lett. {\bf 74}, 3935 (1995).
\bibitem{carl06} B.G. Carlsson and I. Ragnarsson, Phys. Rev. C {\bf 74}, 011302(R) (2006).
\bibitem{beng89} T. Bengtsson, Nucl. Phys. A {\bf 496}, 56 (1989).
\bibitem{beng90} T. Bengtsson, Nucl. Phys. A {\bf 512}, 124 (1990).
\bibitem{Axe02} A. Axelsson, R. Bengtsson and J. Nyberg, Nucl. Phys. A {\bf 708}, 226 (2002). 
\bibitem{rou15} D. G. Roux, W. C. Ma, G. B. Hagemann, H. Amro, D. R. Elema, P. Fallon, A. G\"{o}rgen, B. Herskind, H. H\"{u}bel, Y. Li, {\it et al.}, 
Phys. Rev. C {\bf 92}, 064313 (2015).
\bibitem{Rag81} I. Ragnarssson, S. {\AA}berg and R.K. Sheline, Phys. Scr. {\bf 24}, 215 (1981). 
\bibitem{Sve01} C.E.\ Svensson, A.O.\ Macchiavelli, A.\ Juodagalvis, A.\ Poves, 
I.\ Ragnarsson, S.\ {\AA}berg, D.E.\ Appelbe, R.A.E.\ Austin, 
G.C.\ Ball, M.P.\ Carpenter, {\it et al.}, Phys.\ Rev.\ C {\bf 63}, 061301(R) (2001).
\bibitem{ge12} J. Gellanki, D. Rudolph, I. Ragnarsson, L-L. Andersson, C. Andreoiu, M.P. Carpenter, J. Ekman, 
C. Fahlander, E.K. Johansson, A. Kardan, {\it et al.}, Phys. Rev. C {\bf 86}, 034304 (2012).
\bibitem{Val05} J.\ J.\ Valiente-Dob\'{o}n, T.\ Steinhardt,
C.\ E.\ Svensson, A.\ V.\ Afanasjev, I.\ Ragnarsson, C.\ Andreoiu,
R.\ A.\ E.\ Austin, M.\ P.\ Carpenter, D.\ Dashdorj, G.\ de Angelis, {\it et al.},
Phys.\ Rev.\ Lett.\ {\bf 95}, 232501 (2005).
\bibitem{Afa99} A.\ V.\ Afanasjev, D.\ B.\ Fossan, G.\ J.\ Lane and I.\ Ragnarsson,
Phys.\ Rep.\ {\bf 322}, 1 (1999).
\bibitem{Sta01} K.~Starosta, C.J.~Chiara, D.B.~Fossan, T.~Koike, D.R.~LaFosse,  
G.J.~Lane,  J.M.~Sears, J.F.~Smith, A.J.~Boston, P.J.~Nolan, 
{\it et al.}, Phys.\ Rev.\ C {\bf 64}, 014304 (2001).
\bibitem{Rag93} I. Ragnarsson,  Nucl. Phys. A {\bf 557}, 167c (1993). 
\bibitem{Eva06}  A.\ O.\ Evans, E.\ S.\ Paul, J.\ Simpson, M.\ A.\ Riley, D.E.\ Appelbe, D.\ B.\
Campbell, P.\ T.\ W.\ Choy, R.\ M.\ Clark, M.\ Cromaz, P.\ Fallon, {\it et al.}, Phys.\ Rev.\ C {\bf 73}, 064303 (2006).

\bibitem{Sim94} J. Simpson, M.\ A. Riley, S.\ J. Gale, J.\ F. Sharpey-Schafer, M.\ A.\ Bentley,
A.\ M.\ Bruce, R.\ Chapman, R.\ M.\ Clark, S.\ Clarke, J.\ Copnell,  {\it et al.},
Phys. Lett. B {\bf 327}, 187 (1994).

\bibitem{carl08} B. G. Carlsson, I. Ragnarsson, R. Bengtsson, E. O. Lieder, R. M. Lieder, and A. A. Pasternak, Phys. Rev. C {\bf 78}, 034316 (2008).
\bibitem{ma14} Hai-Liang Ma, B. G. Carlsson, Ingemar Ragnarsson, and Hans Ryde, Phys. Rev. C {\bf 90}, 014316 (2014).

\bibitem{bri05} P. Bringel G.B. Hagemann, H. H\"{u}bel, A. Al-khatib, P. Bednarczyk, A. B\"{u}rger, D. Curien,
G. Gangopadhyay, B. Herskind, D.R. Jensen, {\it et al.}, Eur. Phys. J. A {\bf24}, 167 (2005). 

\bibitem{Jen02} D.R. Jensen, G.B. Hagemann, I. Hamamoto, S.W. {\O}deg{\aa}rd, M. Bergstr\"{o}m, B. Herskind, G. Sletten, S. T\"{o}rm\"{a}nen, J.N. Wilson, 
P.O. Tj{\o}m {\it et al.}, Nucl.\ Phys.\ A \textbf{703}, 3 (2002).
\bibitem{Sch03} G. Sch\"{o}nwa{\ss}er, H. H\"{u}bel, G.B. Hagemann, P. Bednarzyk, G. Benzoni, A. Bracco, P. Bringel, R. Chapman, 
D. Curien, J. Domscheit, B. Herskind {\it et al.}, Phys.\ Lett. B \textbf{552}, 9 (2003).
\bibitem{amro03} H. Amro, W.C. Ma, G.B. Hagemann, R.M. Diamond, J. Domscheit, P. Fallon, A. G\"{o}rgen, 
B. Herskind, H. H\"{u}bel, D.R. Jensen, {\it et al.}, Phys. Lett. B {\bf 553}, 197 (2003).
\bibitem{bri06} P. Bringel, H. H\"{u}bel, A. Al-Khatib, A. B\"{u}rger, N. Nenoff, A. Neusser-Neffgen, G. Sch\"{o}nwa{\ss}er, 
A. K. Singh, G. B. Hagemann, B. Herskind, {\it et al.}, Phys. Rev. C {\bf 73}, 054314 (2006). 

\bibitem{jen04} D. R. Jensen, G. B. Hagemann, I. Hamamoto, B. Herskind, G. Sletten, J. N. Wilson, S. W. {\O}deg{\aa}rd, K. Spohr, H. H\"{u}bel, 
P. Bringel, {\it et al.}, Eur. Phys. J. A {\bf 19}, 173 (2004).

\bibitem{bri07} P. Bringel, C. Engelhardt, H. H\"{u}bel, A. Neusser-Neffgen, S. W. {\O}deg{\aa}rd, 
G. B. Hagemann, C. R. Hansen, B. Herskind, G. Sletten, M. P. Carpenter, {\it et al.}, Phys. Rev. C {\bf 75}, 044306 (2007). 

\bibitem{sch04} G. Sch\"{o}nwa{\ss}er, N. Nenoff, H. H\"{u}bel, G. B. Hagemann, P. Bednarczyk, G. Benzoni, A. Bracco, P. Bringel, R. Chapman, 
D.Curien {\it et al.}, Nucl. Phys. A {\bf 735}, 393 (2004).

\bibitem{mar13} J.C. Marsh, W. C. Ma, G. B. Hagemann, R. V. F. Janssens, R. Bengtsson, H. Ryde, M. P. Carpenter, G. G\"{u}rdal, 
D. J. Hartley, C. R. Hoffman {\it et al.}, Phys. Rev. C {\bf 88}, 041306(R) (2013). 

\bibitem{yad09} R.B. Yadav, W. C. Ma, G. B. Hagemann, H. Amro, A. Bracco, M. P. Carpenter, J. Domscheit, S. Frattini, 
D. J. Hartley, B. Herskind, {\it et al.}, Phys. Rev. C {\bf 80}, 064306 (2009). 

\bibitem{Sch95} H.\ Schnack-Petersen, R.\ Bengtsson, R.\ A.\ Bark, P.\ Bosetti, A.\ Brockstedt, H.\ Carlsson, L.\ P.\ Ekstr\"{o}m, 
G.B.\ Hagemann, B.\ Herskind, F.\ Ingebretsen, {\it et al.}, Nucl.\ Phys.\ A \textbf{594}, 175 (1995).
\bibitem{Ben04} R.\ Bengtsson and H.\ Ryde, Eur.\ Phys.\ J.\ A \textbf{22}, 355 (2004).
\bibitem{Car07} B.G. Carlsson, Int. J. Mod. Phys. E \textbf{16}, 634 (2007).
\bibitem{kar12} A. Kardan, I. Ragnarsson, H. Miri-Hakimabad, and L. Rafat-Motevali, Phys. Rev. C \textbf{86}, 014309 (2012).
\bibitem{Ham03} I. Hamamoto and G.B. Hagemann, Phys. Rev. C \textbf{67}, 014319 (2003).
\bibitem{Shi08} Y.R. Shimizu, T. Shoji and M. Matsuzaki, Phys. Rev. C \textbf{77}, 024319 (2008).
\bibitem{Sug10} K. Sugawara-Tanabe and K. Tanabe, Phys. Rev. C \textbf{82}, 051303(R) (2010).
\bibitem{Alm11} D. Almehed, F. D\"{o}nau and S. Frauendorf, Phys. Rev. C \textbf{83}, 054308 (2011).
\bibitem{Fra14} S. Frauendorf and  F. D\"{o}nau,  Phys. Rev. C \textbf{92}, 064306 (2015).
\bibitem{Rag17} I. Ragnarsson, Phys. Scr. {\bf 92}, 124004 (2017).
\bibitem{Wad11} R. Wadsworth, I. Ragnarsson, B.G. Carlsson, Hai-Liang Ma, P.J. Davies, C. Andreoiu, R.A.E. Austin, 
M.P. Carpenter, D. Dashdorj, S.J. Freeman, {\it et al.}, Phys. Lett. B {\bf 701}, 306 (2011).
\bibitem{Nil95} S. G. Nilsson and I.  Ragnarsson, {\it Shapes and
Shells in Nuclear Structure}  (Cambridge University Press, Cambridge, England, 1995).
\bibitem{strut} V. Strutinsky, Nucl. Phys. A {\bf 95}, 420 (1967).

\bibitem{pomo03} K. Pomorski and J. Dudek, Phys. Rev. C {\bf 67}, 044316 (2003).
\bibitem{rag15} I.\ Ragnarsson, B.G.\ Carlsson, A.\ Kardan and Hailiang Ma, Acta Phys. Polo. B {\bf 46}, 477 (2015).

\bibitem{Juo00} A. Juodagalvis, I. Ragnarsson, S. {\AA}berg, Phys. Lett. B {\bf 477}, 66 (2000).

\bibitem{Wan11} X.\ Wang, M.\ A.\ Riley, J.\ Simpson, E.\ S.\ Paul, J.\ Ollier, R.\ V.\ F.\ Janssens, A.\ D.\ Ayangeakaa, H.\ C.\ Boston, 
M.\ P.\ Carpenter, C.\ J.\ Chiara, {\it et al.}, Phys. Lett. B {\bf 702}, 127 (2011).

\bibitem{jen81} R.\ V.\ F.\ Janssens, M.\ J.\ A.\ de Voigt, H.\ Sakai, H.\ J.\ M.\ Aarts, 
C.\ J.\ Van der Poel, H.\ F.\ R.\ Arciszewski, D.\ E.\ C.\ Scherpenzeel, and J.\ Vervier, Phys. Lett. B {\bf 106}, 475 (1981).

\bibitem{Yu90} C.-H. Yu, G.B. Hagemann, J.M. Espino, K. Furuno, J.D. Garrett, R. Chapman, D. Clarke, F. Khazaie, J.C. Lisle, J.N. Mo,  
{\it et al.}, Nucl. Phys. A {\bf 511}, 157 (1990).

\bibitem{Ben85} T. Bengtsson and I. Ragnarsson, Phys. Lett. B {\bf 163}, 31 (1985).

\bibitem{oga93} S.\ Ogaza, J.\ Kownacki, M.\ P.\ Carpenter, J.\ Gascon, G.\ B.\ Hagemann, Y.\ Iwata, H.\ J.\ Jensen, 
T.\ Komatsubara, J.\ Nyberg, G.\ Sletten, {\it et al.}, Nucl. Phys. A {\bf 559}, 100 (1993).
\bibitem{hart06} D.\ J.\ Hartley, W.\ H.\ Mohr, J.\ R.\ Vanhoy, M.\ A.\ Riley, A.\ Aguilar, C.\ Teal, R.\ V.\ F.\ Janssens, 
M.\ P.\ Carpenter, A.\ A.\ Hecht, T.\ Lauritsen, {\it et al.}, Phys. Rev. C {\bf 74}, 054314 (2006).
\bibitem{Jen01} H.\ J.\ Jensen, R.\ A.\ Bark, P.\ O.\ Tj{\o}m, G.\ B.\ Hagemann, I.\ G.\ Bearden, H.\ Carlsson, S.\ Leoni, T.\ L\"{o}nnrot,
W.\ Reviol, L.\ L.\ Riedinger, {\it et al.}, Nucl. Phys. A {\bf 695}, 3 (2001).
\bibitem{Roh19} A. Rohilla, R.P. Singh, S. Muralithar, A. Kumar, I.M. Govil
and S.K. Chamoli,  Phys. Rev. C \textbf{100}, 024325 (2019).
\bibitem{Gur05} H. G\"{u}rdal, H. Amro, C.W. Beausang, D.S. Brenner, M.P. Carpenter, R.F. Casten, 
C. Engelhardt, G.B. Hagemann, C.R. Hansen, D.J. Hartley, {\it et al.}, J. Phys. G: Nucl. Part. Phys. {\bf 31}, S1873 (2005).

\bibitem{tah18} H.\ Taheri, A.\ Kardan and M.\ H.\ Hadizadeh Yazdi, Phys. Rev. C {\bf 98}, 054313 (2018).

\bibitem{Har65} S.\ M.\ Harris, Phys. Rev. {\bf 138}, B509 (1965).




%
%
%




\end{thebibliography}
\end{document}